\documentclass[twocolumn, astrosymb]{aastex7}
\usepackage{gensymb}
\usepackage{enumitem}
\usepackage{amsmath}

\newcommand{\jykms}{Jy km s$^{-1}$}
\newcommand{\jykmsbeam}{Jy km s$^{-1}$ beam$^{-1}$}
\newcommand{\kms}{km s$^{-1}$}
\newcommand{\halpha}{H$\alpha$}

\newcommand{\DiskWindAngle}{$\phi_\text{DW}=75_{-27}^{+10}$}
\newcommand{\Barolo}{$^\mathrm{3D}\textsc{Barolo}$}

\begin{document}

\title{ALMA-JELLY I: High Resolution CO(2-1) Observations of Ongoing Ram Pressure Stripping in NGC 4858 Reveal Asymmetrical Gas Tail Formation and Fallback  \footnote{Released on XXXX}}

\author[0000-0001-5079-1865]{Harrison J. Souchereau}
\affiliation{Department of Astronomy, Yale University, New Haven, CT 06511}
\email{harrison.souchereau@yale.edu}

\author[0000-0003-0586-6754]{Jeffrey D.P. Kenney}
\affiliation{Department of Astronomy, Yale University, New Haven, CT 06511}
\email{jeff.kenney@yale.edu}

\author[0000-0002-1640-5657]{Pavel J{\'a}chym}
\affiliation{Astronomical Institute of the Czech Academy of Sciences, Boční II 1401, 141 00, Prague, Czech Republic}
\email{jachym@ig.cas.cz}

\author[0000-0001-5880-0703]{Ming Sun}
\affiliation{Department of Physics and Astronomy, University of Alabama in Huntsville, 301 Sparkman Dr NW, Huntsville, AL 35899, USA}
\email{ms0071@uah.edu}

\author[0000-0003-0289-2674]{William J. Cramer}
\affiliation{Department of Physics and Astronomy, Notre Dame University, South Bend, IN 46617, USA}
\email{wcramer2@nd.edu}

\author[0000-0001-7550-2281]{Masafumi Yagi}
\affiliation{National Astronomical Observatory of Japan, Osawa 2-21-1, Mitaka, Tokyo 181-8588 Japan}
\email{YAGI.Masafumi@nao.ac.jp}

\author[0000-0002-9795-6433]{Alessandro Boselli}
\affiliation{Aix-Marseille Univ., CNRS, CNES, LAM, Marseille, France}
\email{alessandro.boselli@lam.fr}

\author[0000-0002-7758-9699]{Elias Brinks}
\affiliation{Centre for Astrophysics Research, University of Hertfordshire, College Lane, Hatfield AL10 9AB, United Kingdom}
\email{e.brinks@herts.ac.uk}

\author[0000-0003-2658-7893]{Francoise Combes}
\affiliation{Observatoire de Paris, LERMA, College de France, CNRS, PSL Univ, Sorbonne University, UPMC F-75014 Paris, France}
\email{francoise.combes@obspm.fr}

\author[0000-0002-7422-9823]{Luca Cortese}
\affiliation{International Centre for Radio Astronomy Research (ICRAR), University of Western Australia, Crawley, WA 6009, Australia}
\email{luca.cortese@uwa.edu.au}

\author[0000-0002-7898-5490]{Boris Deshev}
\affiliation{Astronomical Institute of the Czech Academy of Sciences, Boční II 1401, 141 00, Prague, Czech Republic}
\affiliation{Tartu Observatory, University of Tartu, Observatooriumi 1, 61602 T\~oravere, Estonia}
\email{boris.deshev@asu.cas.cz}

\author[0000-0002-9043-8764]{Matteo Fossati}
\affiliation{Dipartimento di Fisica ``G. Occhialini'', Universit\`a degli Studi di Milano-Bicocca, Piazza della Scienza 3, I-20126 Milano, Italy}
\affiliation{INAF – Osservatorio Astronomico di Brera, Via Brera 28, I-21021 Milano, Italy}
\email{matteo.fossati@unimib.it}

\author[0000-0003-3471-7459]{Romana Grossová}
\affiliation{Astronomical Institute of the Czech Academy of Sciences, Boční II 1401, 141 00, Prague, Czech Republic}
\email{romana.grossova@gmail.com}

\author[0000-0003-4509-7822]{Rongxin Luo}
\affiliation{School of Physics and Electronic Science, Guizhou Normal University, Guiyang 550001, PR China}
\affiliation{Guizhou Provincial Key Laboratory of Radio Astronomy and Data Processing, Guizhou Normal University, Guiyang 550001, PR China}
\email{rongxinluo217@gmail.com}

\author[0000-0001-6729-2851]{Jan Palouš}
\affiliation{Astronomical Institute of the Czech Academy of Sciences, Boční II 1401, 141 00, Prague, Czech Republic}
\email{palous@ig.cas.cz}

\author[0000-0002-3746-4859]{Tom C. Scott}
\affiliation{Institute of Astrophysics and Space Sciences (IA), Rua das Estrelas, 4150–762 Porto, Portugal}
\email{tom.scott@astro.up.pt}



\begin{abstract}

We present new CO(2-1) observations (resolution $\sim$1" = 460pc) of the Coma cluster jellyfish galaxy NGC 4858 obtained from the ALMA-JELLY large program. 
Analyzing this data alongside complimentary Subaru \halpha{} and HST (F600LP / F350LP) observations, we find numerous structural and kinematic features indicative of the effects from strong, inclined ram pressure, including an asymmetric inner gas tail. 
We estimate a highly-inclined disk-wind angle of $\phi_\text{DW}=75_{-27}^{+10}$. By subtracting a simple circular velocity model, we find (1): gas clumps that are being accelerated by ram pressure, and (2): signatures of gas clumps that had been previously pushed out of the disk but are now falling inwards. 
We also discuss head-tail morphologies in star complexes within the stellar disk that appear to be RPS-influenced. 
Lastly, we compare this galaxy to state-of-the-art galaxy "wind tunnel" simulations. 
We find that this galaxy is one of the best nearby examples of strong and inclined ram pressure gas stripping, and of gas that is perturbed by ram pressure but not fully stripped and falls back. We emphasize the importance of torques due to ram pressure in highly-inclined interactions, which help drive gas inwards on the side rotating against the wind, contributing to the formation of asymmetric inner RPS tails. 

\end{abstract}


\section{Introduction} 
\label{sec:intro}

 Ram pressure affects galaxies traveling through any fluid medium, such as the intracluster medium (ICM) of a galaxy cluster \citep{gunn_infall_1972, boselli_ram_2022}, the intragroup medium of a galaxy group (IGM) \citep{roberts_lotss_2021}, or the circumgalactic medium (CGM) of a larger central host galaxy \citep{geha_slar_2012, samuel_jolt_2023}. The result of sufficiently strong ram pressure can be gas removal from galaxies, called ram pressure stripping (RPS). This process can be an effective environmental quenching mechanism for spiral galaxies \citep{abadi_ram_1999, boselli_environmental_2006, tonnesen_environmentally-driven_2007, chung_vla_2009, wetzel_galaxy_2012, cortese_dawes_2021, jian_radial_2023, brown_vertico_2023}. 
 The effects of RPS are more pronounced for less gravitationally bound and lower surface-density gas, and are seen in HI surveys of cluster galaxies, which have shown that spiral galaxies are deficient on average in HI content, suggesting that the gas has been removed from the disk of the galaxy \citep{giovanelli_gas_1985, cayatte_vla_1990, solanes_h_2001, chung_vla_2009, boselli_spectacular_2016, taylor_faint_2020}. 
 The stripping can manifest as dramatic gas tails extending opposite to the galaxy's direction of motion through the cluster \citep[e.g.][]{abadi_ram_1999, yagi_dozen_2010, yagi_extended_2017, gavazzi_ubiquitous_2018, jachym_alma_2019, cramer_spectacular_2019}. Many of these tails contain young stars, indicating that star formation can occur even after the gas has been stripped \citep[e.g.][]{kenney_ongoing_1999, sun_spectacular_2010, poggianti_gasp_2017, george_ultraviolet_2023, vulcani_evidence_2024}, although star formation in tails is not ubiquitous \citep[e.g.][]{gavazzi_75_2001, boselli_spectacular_2016}. 
 A subset of RPS galaxies are known as "jellyfish galaxies" due to the presence of stars in the tail \citep{ebeling_jellyfish_2014, mcpartland_jellyfish_2016}. 
  

The consequences of a ram pressure event may differ greatly depending on the angle between the wind and the disk \citep{roediger_ram_2006, jachym_ram_2009}. In the case of a highly-inclined wind (closer to edge on), galaxy rotation will influence the resultant ram pressure strength, leading to uneven forces across the galaxy disk. Simulations have shown that, for a similar ram pressure strength, an edge-on wind is less efficient at stripping gas than a face-on wind \citep{jachym_ram_2009}. 
Lower stripping efficiency may elevate fallback, where gas is accelerated out of the gas disk but then falls back inwards. This can lead to a "fountain" effect, where gas is both being stripped and reentering the disk in the inner tail \citep{sparre_magnetised_2023}.
This interaction between rotation and ram pressure can lead to an asymmetric tail morphology, which is observed in galaxy-scale wind-tunnel simulations \citep{vollmer_ram_2001, roediger_star_2014, akerman_how_2023}.  Exploring these effects is important as ram pressure winds are most likely to be highly-inclined.
The effect on gas within a galaxy's disk can also differ based on the disk-wind angle. For highly-inclined events, gas on the leading side (the side closest to the ram pressure wind) will be further pushed into the disk, resulting in compression and potentially increased star formation \citep{koopmann_h_2004, kenney_hst_2015, cramer_alma_2020, troncoso-iribarren_better_2020, boselli_virgo_2021, zhu_when_2023, vulcani_evidence_2024}.

\begin{figure}
    \centering    
    \includegraphics[width=0.49\textwidth]{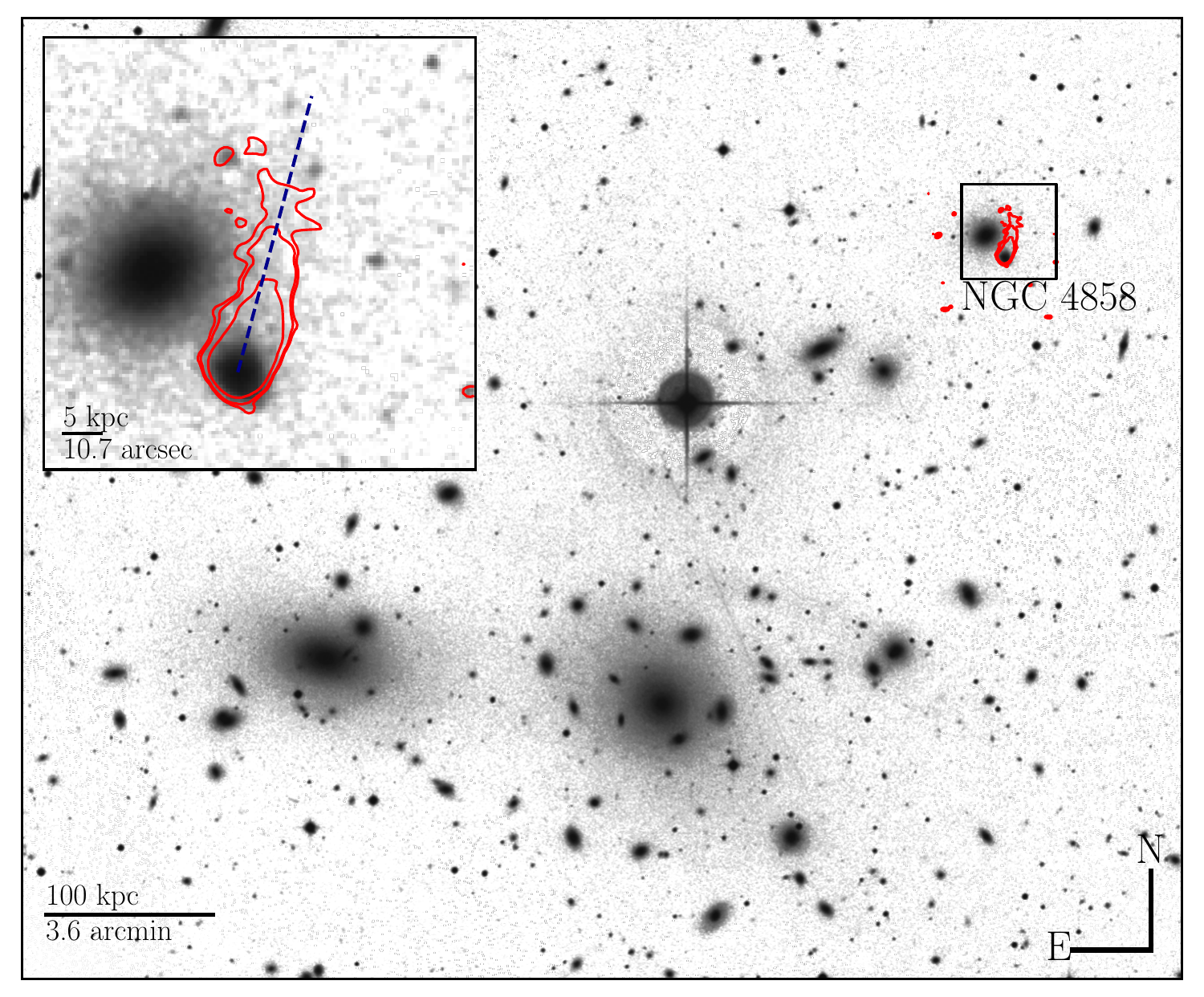}
    \caption{The central region of the Coma cluster, with the location of NGC 4858 indicated, and shown as a zoomed-in cutout. The two BCGs in the center of Coma (NGC 4874 and NGC 4889) can be seen in the center-left of the image. The 144 MHz radio continuum tail (see \citealt{roberts_lotss_2021}) is shown with red contours, and the estimated plane-of-sky tail angle is displayed as the dashed line. \\
    Image from the Digitized Sky Survey (from NASA/ESA \href{https://esahubble.org/images/heic0901c/}{https://esahubble.org/images/heic0901c/}). }
    \label{fig:comacluster}
\end{figure}

The role and importance of fallback in galaxy evolution under ram pressure is not fully understood, but it has been observed in both real galaxies and simulations \citep{vollmer_ram_2001, tonnesen_star_2012, koppen_ram_2018, cramer_molecular_2021, zhu_when_2023, sparre_magnetised_2023}. 
Simulations have shown that a considerable fraction of the gas accelerated above the disk does not reach the escape speed \citep{tonnesen_star_2012}. The recycling nature of fallback can have implications for galaxy evolution, including effects on mass-loss rates and the mixing of cold ISM gas with hot ICM gas.
Molecular gas observations can reveal fallback most effectively, as cold, dense gas is the hardest to strip and most likely to fall back after being perturbed by ram pressure. The relationship between highly-inclined winds, asymmetric inner tails, and fallback is an open question addressed by this work.

\subsection{The ALMA-JELLY Project}
\label{sec:alma-jelly}

In this paper we present the first results from the ALMA-JELLY program. The goal of this program is to map CO(2-1) at 1" resolution in a selection of very strong RPS in nearby high-mass clusters. The ALMA JELLY galaxies reside in the Coma, Leo, and Norma clusters, with 18, 8, and 2 galaxies in each cluster respectively. 

ALMA-JELLY comprises combined 12m and 7m data obtained during ALMA Cycle 8 \footnote{PID 2021.1.01616.L} for 27 of the galaxies, with 1 (ESO 137-001) obtained from archival data \citep[See][]{jachym_alma_2019}. The galaxies were selected based on having a prominent one-sided \halpha{} ionised gas tail in three nearby rich clusters, Norma, Leo and Coma \citep{sun_spectacular_2010, yagi_dozen_2010,fossati_65_2012, yagi_extended_2017, gavazzi_ubiquitous_2018}. The galaxy sample covers a broad range of stellar masses ($\sim10^{7.5} - 10^{10} M_\odot$). It also includes a wide range of evolutionary stage of stripping, from early stages with tails connected to star-forming disks, to late stages with detached tails, and passive or post-starburst disks. The complete sample and full description of the project will be presented in Jachym et al. (in prep.).

 NGC 4858 
 is one of the best nearby examples of a galaxy with strong CO emission and a multi-component asymmetric molecular gas tail. This makes it an ideal galaxy to demonstrate the quality of the ALMA-JELLY data and the science made possible through this program. 
 
 \subsection{The Coma Cluster and NGC 4858}

 \begin{figure*}
    \centering
    \includegraphics[width=\textwidth]{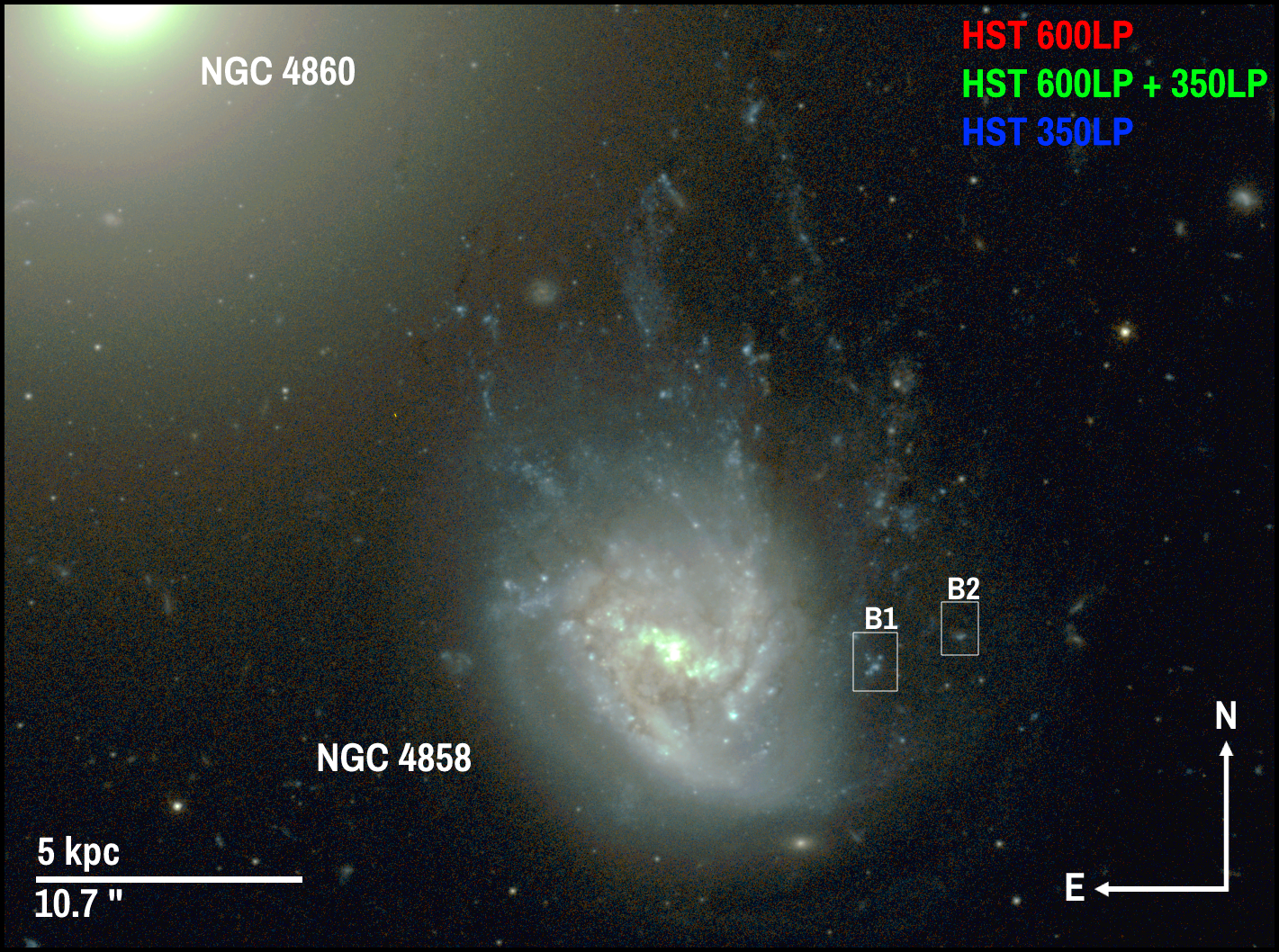}
    \caption{HST F350LP (blue) and F600LP (red) image of NGC 4858. The green channel is an average of the red/blue channels. The background elliptical galaxy NGC 4860 can be seen in the upper left. The stellar distributions coincident with the western \halpha{} blobs, B1 and B2, are indicated.}
    \label{fig:N4858-Zoomin}
\end{figure*}

\begin{deluxetable}{ccccc}[t]
\tablecaption{Properties for \textit{NGC 4858} and (when applicable) its neighbor \textit{NGC4860}. (1,2): Right ascension and declination. (3): B,V, and i-band magnitudes. (4,5): Line-of-sight velocity velocity and line-of-sight velocity relative to the Coma cluster. (6): Distance from the center of the Coma cluster. (6,7,8): Stellar mass, star formation rate measured using NUV and $22\mu m$ luminosity, and HI gas mass estimates from \cite{molnar_westerbork_2022}. (9) HI deficiency factor measured from the M$_{HI}$ - M$_*$ relationships in \cite{parkash_relationships_2018}. (10) $H_2$ gas mass inferred from integrated CO flux. (11) \halpha{} flux. \\ \textbf{References}: A: This work. B: SIMBAD database. C: \cite{molnar_westerbork_2022}. D: \cite{yagi_dozen_2010}  \label{tab:galaxyparams}}
\tablehead{
 & \colhead{Parameter} & \colhead{\textbf{NGC 4858}} & \colhead{NGC4860} & Ref.}
\startdata
   (1) & $\alpha$ (J2000) & 12:59:02.072 & 12:59:03.906 & B\\
   (2) & $\delta$ (J2000) & +28:06:56.228 & +28:07:25.326 & B\\
   (3) & B, V, i [mag] & 15.5 15.0 14.8 & 14.7 13.2 12.7 & B\\
   (4) & Velocity $[cz]$ & 9416 & 7930 &  B\\
   (5) & $V_{rel}$ $[cz]$ & 2421 & 934 & B\\
   (6) & $M_*$ [$10^9M_\odot$] & $4.90 \pm 0.10$ & 44.6 - 91.2 & A,C\\
   (7) & SFR [$M_\odot / \mbox{yr}$] & $3.55 \pm 0.2$ &  &  C\\
   (8) & $M_{HI}$ [$10^9M_\odot$] & 0.25 &  &  C \\
   (9) & Def$_{HI}$  & 0.94 & & C \\
   (10) & $M_{H_2}$ [$10^9M_\odot$] & 2.0 & &  A \\
   (11) & F$_{H_\alpha}$ [erg s$^{-1}$] &  $1.0\times10^{-13}$ & & D \\ 
\enddata
\end{deluxetable}

The Coma cluster's proximity ($\sim100$ Mpc) and high mass $\left(M > 10^{15}M_\odot\right)$ makes it an ideal environment for studying  strong RPS events. The location of NGC 4858 in the cluster, and its tail direction (See Section \ref{sec:tail-asymmetry} for more details) is indicated in Figure \ref{fig:comacluster}. 

NGC 4858 has one of the smallest projected distances to the cluster centre out of the ALMA-JELLY sample, at 355 kpc ($\sim$0.14 $r_{vir}$) \citep{lokas_dark_2003}. The galaxy is highly redshifted with respect to the cluster mean velocity, with a relative velocity of $v_{rad} = +2421$ \kms. The combination of small projected distance and high line-of-sight velocity puts NGC 4858 at the upper envelope of galaxies in the Coma cluster phase-space diagram \citep[See Figure 11 of][]{boselli_ram_2022}, suggesting it has recently entered the cluster. Because the galaxy is redshifted with respect to the Coma cluster mean, the downstream direction of the tail is pointed towards the observer. We estimate the galaxy's disk-wind angle (the angle between the vector normal to the galaxy disk plane and the wind vector) to be $\phi_\text{DW}=75_{-27}^{+10}$ in Section \ref{sec:tail-geometry}.

 NGC 4858 has an HI gas mass of $2.5\times10^8$M$_\odot$ \citep{molnar_westerbork_2022}. This corresponds to an HI deficiency factor of Def$_{HI} = 0.94$, suggesting that NGC 4858 is missing more than 90\% of its expected HI gas \citep{parkash_relationships_2018}. 
NGC 4858 is overall blue in color, indicating significant recent star formation.  \citet{gallego_observations_1996} identify NGC 4858 as a bright \halpha{} emitting galaxy. \citet{yagi_dozen_2010} discovered the presence of a large \halpha{} gas tail extending north from the galaxy disk, indicating some combination of shock-excited gas and/or star-formation in the tail. The distribution of \halpha{} is highly asymmetric, with more emission on the northwest side of the galaxy, which is the trailing side of the galaxy rotating into the wind.

NGC 4858's disk and tail are bright in NUV/FUV emission  \citep{smith_ultraviolet_2010}. The alignment of UV and \halpha{} emission, combined with the presence of bright \halpha{} clumps in the tail, suggests that a significant fraction of the \halpha{} is coming from HII regions around newly-formed stars. This is further confirmed by  the HST F350LP/F600LP image of the galaxy (see Figure \ref{fig:N4858-Zoomin}) which shows clumps of blue stars throughout the tail. The presence of stars in the tail makes NGC 4858 a ``jellyfish galaxy".

 The galaxy has an X-ray tail observed in 0.5-2 keV with Chandra, indicating the presence of hot gas in the tail \citep[See][supplementary Figure 11]{sun_universal_2021}. 
 The galaxy also has a radio continuum tail that extends approximately 30-40 kpc north of the galaxy disk, first revealed in \cite{chen_ram_2020} and in \cite{roberts_lotss_2021, roberts_radio-continuum_2024}. 
 The large physical extent of this tail provides us with the most reliable estimate for the on-sky tail angle for the galaxy, which we estimate as $\theta_\text{tail} = -15\pm20\degree$ measured counterclockwise from north (See Section \ref{sec:tail-geometry}).




The elliptical galaxy NGC 4860 (GMP 3792) is observed nearby. Although these galaxies appear close in projection ($\sim 40"$), their high relative velocities (approximately 1500 \kms) along the line of sight, combined with no evidence for tidal features in either galaxy, suggests that they are not interacting. 


The Coma cluster has a redshift of $z=0.0235$ and a velocity dispersion of $\sim 950$ \kms \citep{sohn_velocity_2017}. We will use the standard cosmological parameters from \cite{hinshaw_nine-year_2013} ($H_0=69.7$ \kms{} Mpc$^{-1}$). 
We adopt a luminosity distance for Coma of 100 Mpc, corresponding to a scale of 1" = 0.463 kpc. 
NGC 4858 has an estimated stellar mass of $4.9\times10^9 M_\odot$, making it an intermediate mass spiral galaxy \citep{molnar_westerbork_2022}.


\subsection{Outline}


We describe in Section \ref{sec:data} the new ALMA CO data and ancillary datasets. After analyzing the stellar body (Section \ref{sec:stellar-distribution}), we explore the CO distribution and compare to other tracers (Section \ref{sec:co-structure}). We then explore the global CO kinematics (Section \ref{sec:co-kinematics}) and describe individual regions with highly non-circular motions (Section \ref{sec:high-residuals}). In Section \ref{sec:asymmetries} we discuss the theory behind the development of asymmetries in the inner RPS tail. and compare the observational data to simulations. We summarize the results in Section \ref{sec:conclusion}. We include a detailed discussion of deriving the disk-wind angle in Appendix \ref{app:disk-wind-angle}, and describe in detail the possible origins for observed velocity residuals in Appendix \ref{app:high-velocity-residuals}. 

\section{Data and Data Reduction}
\label{sec:data}

\begin{figure*}
    \centering
    \includegraphics[width=0.9\textwidth]{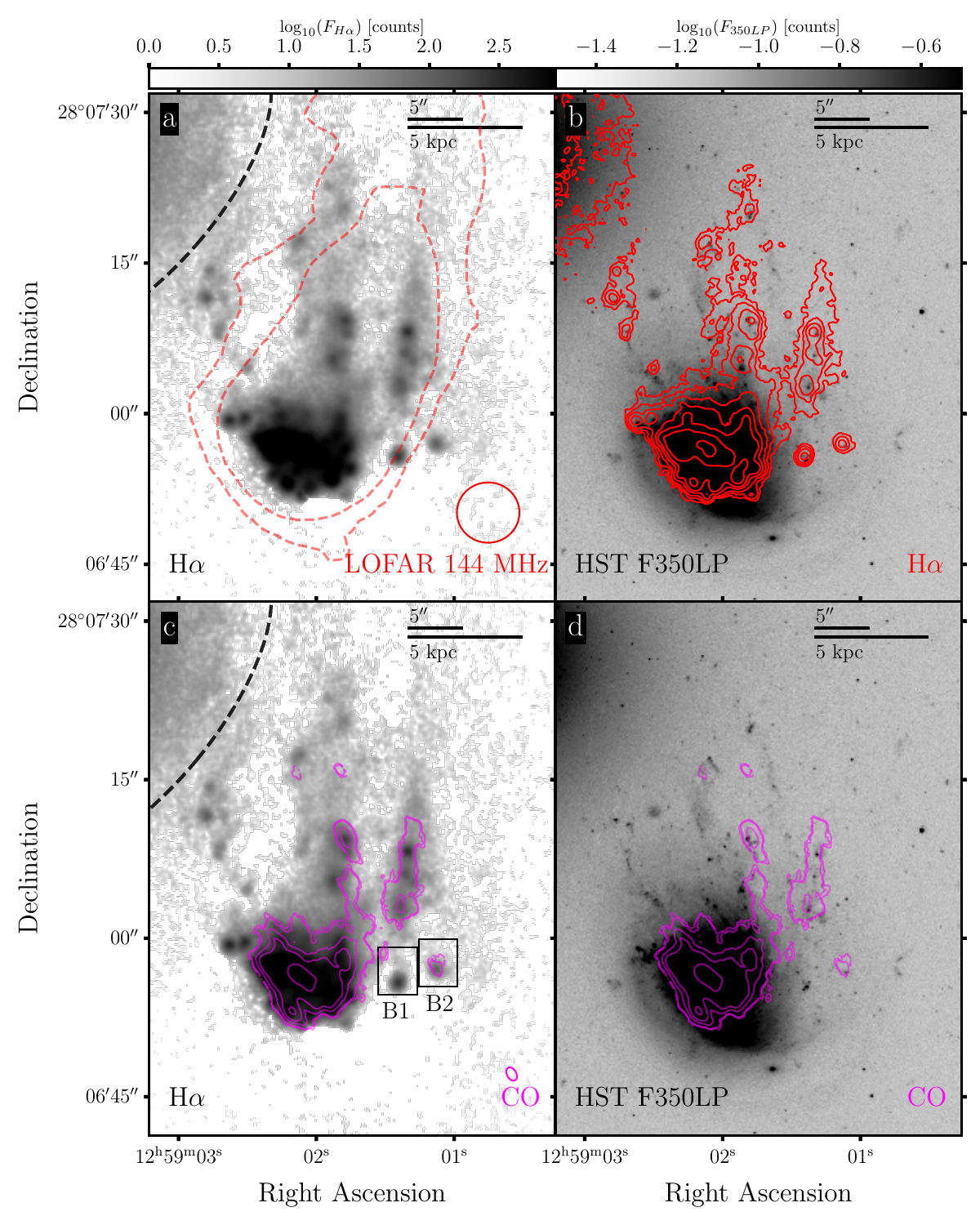}
    \caption{Overlay maps for NGC 4858. From the top left: (a): The Subaru \halpha{} distribution, with contours showing the LOFAR 144 MHz radio continuum tail in red. (b): The HST 350LP image in greyscale with \halpha{} contours in red. (c): The Subaru \halpha{} image with ALMA CO(2-1) contours in magenta. (d): The HST greyscale image with the same ALMA CO contours. The CO contours  (in multiples of the noise level $\sigma_\text{mom0} = 0.031$\jykmsbeam) are [2, 3, 10, 20, 50]. For \halpha, the contours are [5, 10, 20, 50, 100, 250, 1000] multiples of the \halpha{} noise level of 3.7 counts. The beams for the LOFAR data and ALMA CO(2-1) are indicated in the left column. 
    The black dashed lines in panels (a) and (c) indicate the region near the elliptical galaxy likely to be affected by imperfect continuum subtraction.}
    \label{fig:overlay-maps}
\end{figure*}

Alongside the new ALMA CO(2-1) data that we present in this work, we have supplementary Subaru \halpha{} and deep HST 350LP and 600LP photometry, allowing us also to look at the distributions of excited gas and young stars. We use the astropy subpackage \textsc{Reproject} \footnote{See: \href{https://reproject.readthedocs.io/en/stable/\#}{https://reproject.readthedocs.io/en/stable/\#}} to align and rescale the images to the ALMA CO coordinate frame. To show how our different wavelength tracers compare with each other, we present overlay maps in Figure \ref{fig:overlay-maps}.

\subsection{ALMA CO(2-1)}
\label{sec:alma-co}

NGC 4858 was observed by ALMA with 5 mosaic pointings using the 7m array, and 13 mosaic pointings with the 12m array.
ALMA observed the 230.5 GHz CO(2-1)line, and any subsequent mention of CO emission implies the CO(2-1) transition unless stated otherwise. 

The ALMA data were cleaned using the PHANGS-ALMA data reduction pipeline \citep{leroy_phangsalma_2021}, which we will briefly describe. Instead of calling \textsc{tclean} once to clean a dataset, as is the standard procedure, the PHANGS-ALMA pipeline differs by calling \textsc{tclean} multiple times over multiple scales, while supplying masks to guide deconvolution. The data is also corrected for the primary beam, and then automatically trimmed to an appropriate image size. We choose to use a natural weighting scheme (Briggs \texttt{robust=2}) to give us the best surface brightness sensitivity.

The datacube has 480 x 303 pixels (corresponding to $72\times45$ arcseconds, or  $34 \times 21$ kpc, respectively), and channel widths of 3.8 \kms. The synthesized primary beam has major and minor axes of 1.35 and 0.87 arcseconds, respectively, and a position angle of $30\degree$. The RMS of the cube is measured at $\sim 2.3$ mJy Beam$^{-1}$.

Moment maps (CO emission, mean velocity, and velocity dispersion) from the masked datacube are shown in Figure \ref{fig:momentmaps}. We generate masks for the cleaned datacube using the Python \textsc{maskmoment} package, developed by Tony Wong \footnote{See \href{https://github.com/tonywong94/maskmoment}{https://github.com/tonywong94/maskmoment}}. 
This maximizes the amount of geniune emission included in the masked cube. The weakest features are required to be detected at $>3.0\sigma$ in at least 2 contiguous channels, and must cover a minimum spatial area as large as the synthesized beam.




From the moment-0 map, (Figure \ref{fig:momentmaps}), we measure an integrated CO flux \textbf{(disk + tail)} of $S_{CO}\Delta v = 46.2 \pm 2.7$ \jykms. We can then follow the procedure in \cite{solomon_molecular_1997} to obtain a CO line luminosity (in units of brightness temperature) using

\begin{equation}
    L'_\text{CO} = 3.25\times10^7 \left(S_{CO}\Delta v\right) \nu_\text{obs}^{-2}D_L^2 (1 + z)^{-3}
\end{equation}

\noindent where $\nu_\text{obs}$ is the observed frequency of the CO(2-1) line in GHz, and $D_L$ is the luminosity distance in Mpc (100 Mpc at $z=0.0231$).  We then convert the CO luminosity to a total H$_2$ mass using

\begin{equation}
    M_{\text{H}_2} = \frac{\alpha_{CO}}{R_{21}}L'_\text{CO}
\end{equation}

\noindent where $R_{21} \approx 0.8$ \citep{leroy_heracles_2009} is the CO(2-1)/CO(1-0) ratio, and $\alpha_{CO} = 4.3\text{M}_\odot$ (K \kms{} pc$^2$)$^{-1}$ \citep{bolatto_co--h2_2013}. 
Using this conversion, we estimate a molecular hydrogen mass of M$_{\text{H}_2} = 1.5\pm0.1\times10^9M_\odot$. Dividing this by the estimated stellar mass for NGC 4858 ($4.90\times10^9 M_\odot$, \citealt{molnar_westerbork_2022}) gives us our $M_{H_2}$/$M_*$ ratio:

\begin{equation}
    \frac{M_{H_2}}{M_*} = \frac{1.5\times10^9 M_\odot}{4.9\times10^9 M_\odot} = 0.31
\end{equation}

This ratio is high, given that field galaxies at the stellar mass of NGC 4858 have typical ratios of $M_{\text{H}_2}/M_* \approx 0.08$ \citep{popping_evolution_2014}. 
The large H$_2$ mass fraction is likely due to a combination of an intrinsically high mass fraction, plus the adopted Milky Way value for CO/$M_{H_2}$ possibly being too high for RPS galaxies. Since the L$_{CO}$/M$_*$ is high, the large HI deficiency factor in NGC 4858 might be caused in part by HI being converted to H$_2$ \citep{moretti_high_2020}.

\subsection{Suprime Cam}

Ground-based R-band and narrow-band (\halpha{} at the Coma cluster) images of NGC 4858 come from \citet{yagi_dozen_2010}, to which we refer the reader for a more in-depth explanation of the data acquisition and processing \citep[also][]{yagi_extended_2017}. 
These images were obtained using the Suprime-Cam, located at the prime focus of the Subaru telescope \citep{miyazaki_subaru_2002}.

In the northeast region near the elliptical galaxy, continuum subtraction for the \halpha{} image is imperfect and not all continuum starlight is removed from the narrow-band filter. We indicate this region in Figure \ref{fig:overlay-maps} (panels a and c). 
This is unlikely to affect analysis of the \halpha{} tail of NGC 4858, where the continuum is low and the distribution is not smooth.

\subsection{HST Imaging}

For this project we also use publicly-available HST Wide-Field Camera 3 (WFC3) F350LP and F600LP images from the \textit{HST Legacy Archive} (PI Gregg, Proposal ID: 13777), shown in Figure \ref{fig:N4858-Zoomin}. The wavelength bands correspond to $3400$\AA{} and $5800$\AA{} rest-frame emission for NGC4858, respectively. The 350LP image best reveals both the distributions of young stars in the galaxy, as well as dust extinction. The data described here may be obtained from the MAST archive at
\dataset[doi:10.17909/ts8f-7f96]{https://dx.doi.org/10.17909/ts8f-7f96}.


\subsection{LOFAR Imaging}

The LOFAR 144 MHz radio data used in this work come from the Lofar Two-Meter Sky Survey (LoTSS) \citep[See][]{shimwell_lofar_2017}. 
A cutout surrounding emission from NGC 4858 was created from the larger Coma cluster mosaic image \citep[For a more detailed analysis of this region, see][]{roberts_lotss_2021}. 
From this cutout we estimate a total 144 MHz flux density of $5\times10^{-4}$ Jy for NGC 4858.

\begin{figure*}
    \centering
    \includegraphics[width=\textwidth]{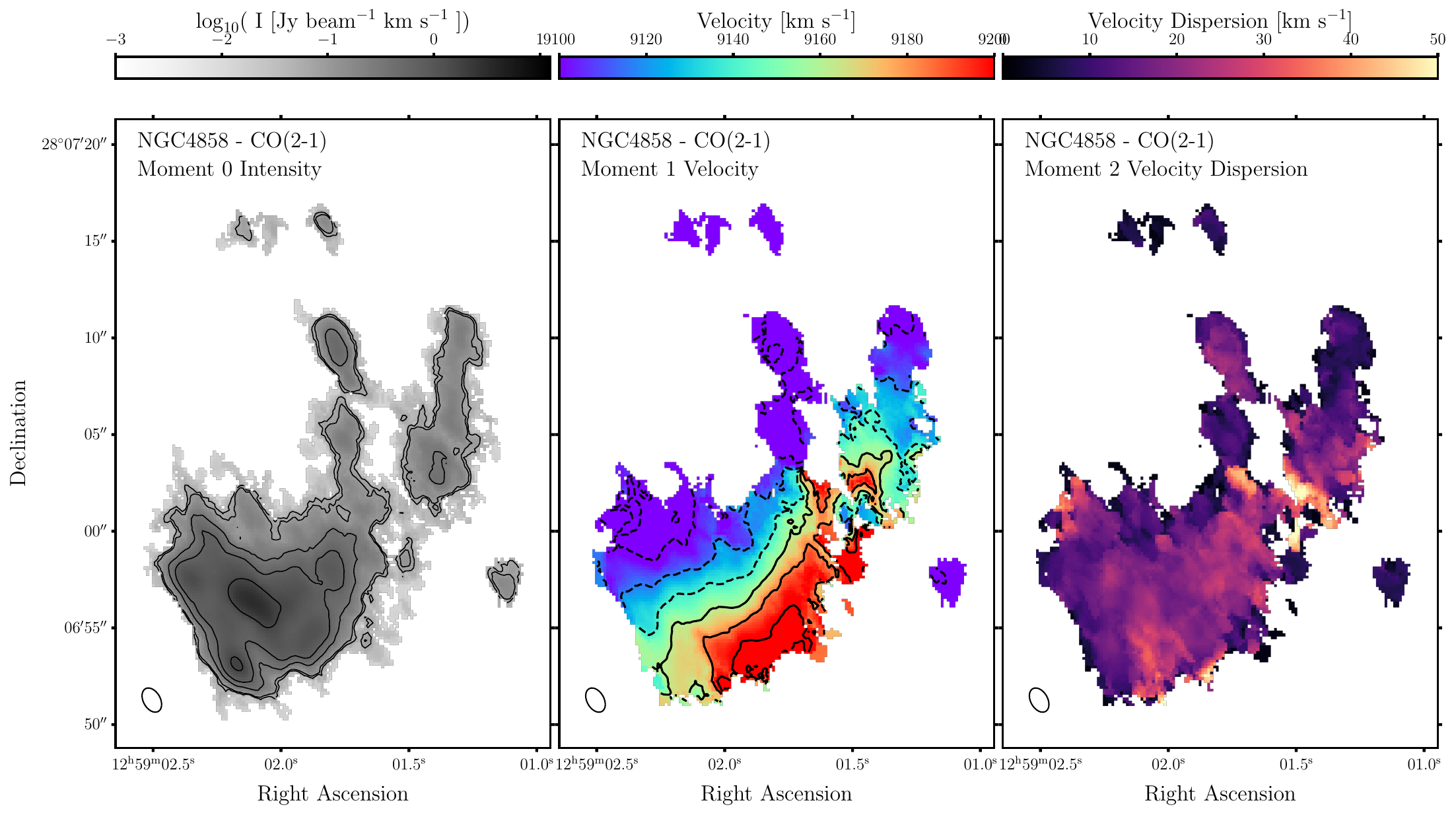}
    \caption{CO(2-1) moment maps of NGC 4858. \textbf{Left:} Moment-0 with overlaid isophotal contours to show structure. 
    The isophotal contour levels are the same as in Figure \ref{fig:overlay-maps}.
    \textbf{Center:} Moment 1 with overlaid isovelocity contours (dashed lines indicate blue-shifted velocities).  
    \textbf{Right:} Moment 2 map. The beam size ($1.35" \times 0.87"$) is shown in the bottom left for all three panels.}
    \label{fig:momentmaps}
\end{figure*}

\section{Stellar Distribution}
\label{sec:stellar-distribution}

\begin{figure*}
    \centering
    \includegraphics[width=\textwidth]{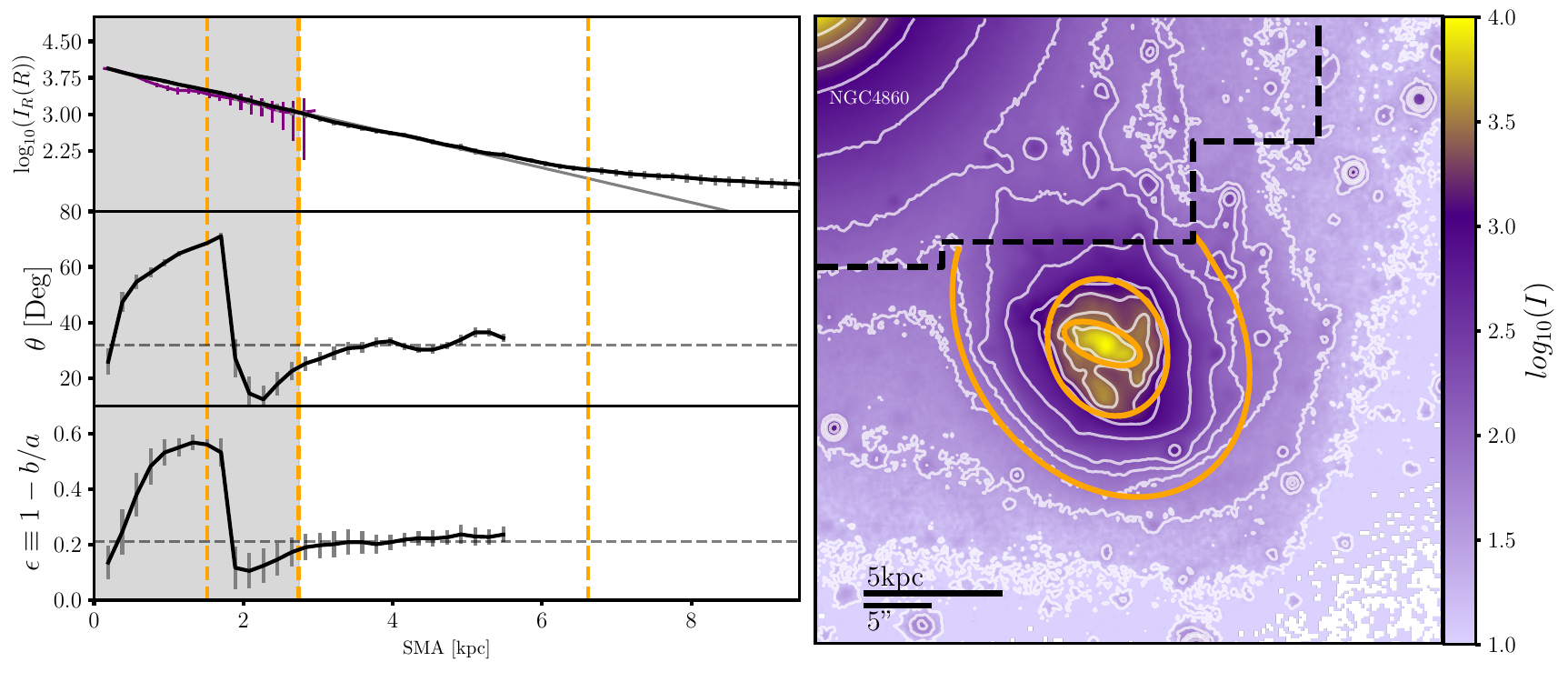}
    \caption{Isophotal analysis of the Subaru R-band image of NGC 4858. \textbf{Left:} From top to bottom: surface brightness (and S{\'e}rsic profile fit in grey), position angle measured from North in degrees, and ellipticity ($1 - b/a$), against semimajor axis in kpc. The radial profile of the CO distribution out to the truncation radius is shown in purple. The dashed orange lines indicate, in order of increasing radii: the extent of the bar-dominated region, the extent of the truncated side of the CO gas disk (also highlighted by the grey shaded region), and the radial extent to which the starlight is dominated by that of NGC 4858's stellar disk. \textbf{Right:} The cutout used in isophotal analysis. The masked region is indicated by the dashed black line, North of which all pixels (contaminated by NGC4860) are masked. Isophotes are plotted in white. The orange ellipses correspond to the same dashed lines in the left panels.}
    \label{fig:isophotalanalysis}
\end{figure*}


In order to properly interpret the gas distribution and kinematics of NGC 4858, it is important to characterize the stellar body of the galaxy. The stellar disk of NGC 4858 (see Figure \ref{fig:N4858-Zoomin}) has a prominent bar and two spiral arms apparently originating from its ends. Based on the spiral arm orientation, we infer that the galaxy is rotating in a clockwise direction \citep{iye_spin_2019}. 

To characterize the distribution of old stars in NGC 4858, we use the Suprime-Cam R-band image, which is deeper than the HST image. We use the elliptical isophote analysis tool in \textsc{GalPRIME}, which has a modified wrapper for the \textsc{Photutils} algorithm \footnote{\href{https://github.com/HSouch/GalPRIME}{https://github.com/HSouch/GalPRIME}} (Souchereau et al., in prep.). This method is an iterative process that fits each isophote individually for $(x,y)$ centre, position angle, ellipticity, and upper harmonic coefficients, such as $a_3$ and $a_4$ \citep{jedrzejewski_ccd_1987}.

Extended starlight from NGC 4860 contaminates the region to the North. We mitigate this contamination by heavily masking the Northern region of the image, which in isophotal analysis plots is denoted by a thick dotted line. Although the light from the elliptical galaxy does extend beyond this masked area and combines with the light distribution from NGC 4858, the proportion of light from the elliptical galaxy in the unmasked regions is so small that it does not significantly affect the parameters derived from the unmasked light distribution.

The results of the isophote fitting on the R-band image are shown in Figure \ref{fig:isophotalanalysis}. At approximately 2kpc, there is a significant change in both the position angle and ellipticity of fit isophotes. This abrupt change coincides with the radius where the stellar light distribution changes from bar-dominated to disk-dominated. Beyond the CO truncation radius ($\sim$ 2.7 kpc) to where the algorithm stops fitting individual isophotes ($\sim 5.7$ kpc) the position angle and ellipticity ($\epsilon = 1 - b/a$) are stable. We estimate a position angle of $\theta_\text{maj} = 32 \pm 6\deg$ and an ellipticity of $\epsilon = 0.21 \pm 0.05$ respectively. This corresponds to an inclination of $38\degree$.

We fit a 1D S{\'e}rsic profile model to the light distribution of NGC 4858. From this profile fit, we infer an effective radius of $r_e = 4.4" = 2.1$kpc (corresponding to an exponential scale length of 1.26 kpc) and a S{\'e}rsic index of $n=0.98$.  Because NGC 4858 is well fit by an $n=1$ S{\'e}rsic profile out to $R = 15.1" = 7.05$kpc, this is an indication of no significant starlight contribution from a stellar bulge. 

The right panel of Figure \ref{fig:isophotalanalysis} shows isophotal contours compared to elliptical isophotes from the profile extraction. The isophote corresponding to the radial distance where the extracted profile deviates from the S{\'e}rsic model fit is indicated as the largest orange ellipse. This also corresponds with isophotal contours that begin to more directly trace light from the nearby elliptical galaxy, so we cannot reliably measure the light profile of the stellar disk any further. This distance of 7kpc is our lower limit on the size of the stellar disk.



\section{CO Distribution \& Comparisons with Other Wavelengths}
\label{sec:co-structure}
  
The CO moment-0 map (See Figure \ref{fig:momentmaps}) traces the molecular gas distribution in NGC 4858.
Within the disk, the galaxy has some relatively normal features, such as a prominent CO bar aligned with the stellar bar ($\theta_\text{bar,CO} \approx \theta_\text{bar,*} = 70\deg$), as well as CO peaks coincident with the stellar spiral arms. Other features, such as along the southern leading side, suggest that the galaxy is experiencing ongoing strong and highly-inclined ram pressure.

Gas beyond 3kpc from the galaxy center is highly asymmetric, with a molecular tail to the north and truncation along the southern side, which is the side of the galaxy most exposed to the ram pressure wind. The majority of CO flux ($\sim 86\%$ of the total flux) is located within the leading-side CO truncation radius.

In the northern, downstream direction, CO emission is detected out at least as far as $21" = 10$kpc. 14\% of the total CO emission is located beyond the southern side truncation radius, most of it in 2 prominent arms that both extend northwards. We refer to this region as the inner tail region, although some of the gas may be in or near the disk. Some of the gas in this region may be disturbed but not be sufficiently accelerated by ram pressure to escape the galaxy. This inner tail, and the behavior of gas in this region, is a central focus of this paper.


Using the same methods as for the R-band stellar distribution, we extract a radial surface brightness profile from the moment-0 map. We find that the CO gas within the truncation radius is well fit by an exponential disk profile with an exponential scale length of  $h_R = 2.49" = 1.16$kpc, about the same as the stellar disk.  
The extent of the CO gas is only approximately 1.9 CO scale lengths (2.7 kpc) along the southern side. The regularity of the CO gas disk suggests that the remaining gas disk is not strongly disturbed by ram pressure, except at the edges of disk where the wind is most strongly felt. 


\subsection{Tail Asymmetry}
\label{sec:tail-asymmetry}

\begin{figure*}
    \centering
    
    \includegraphics[width=0.60\textwidth]{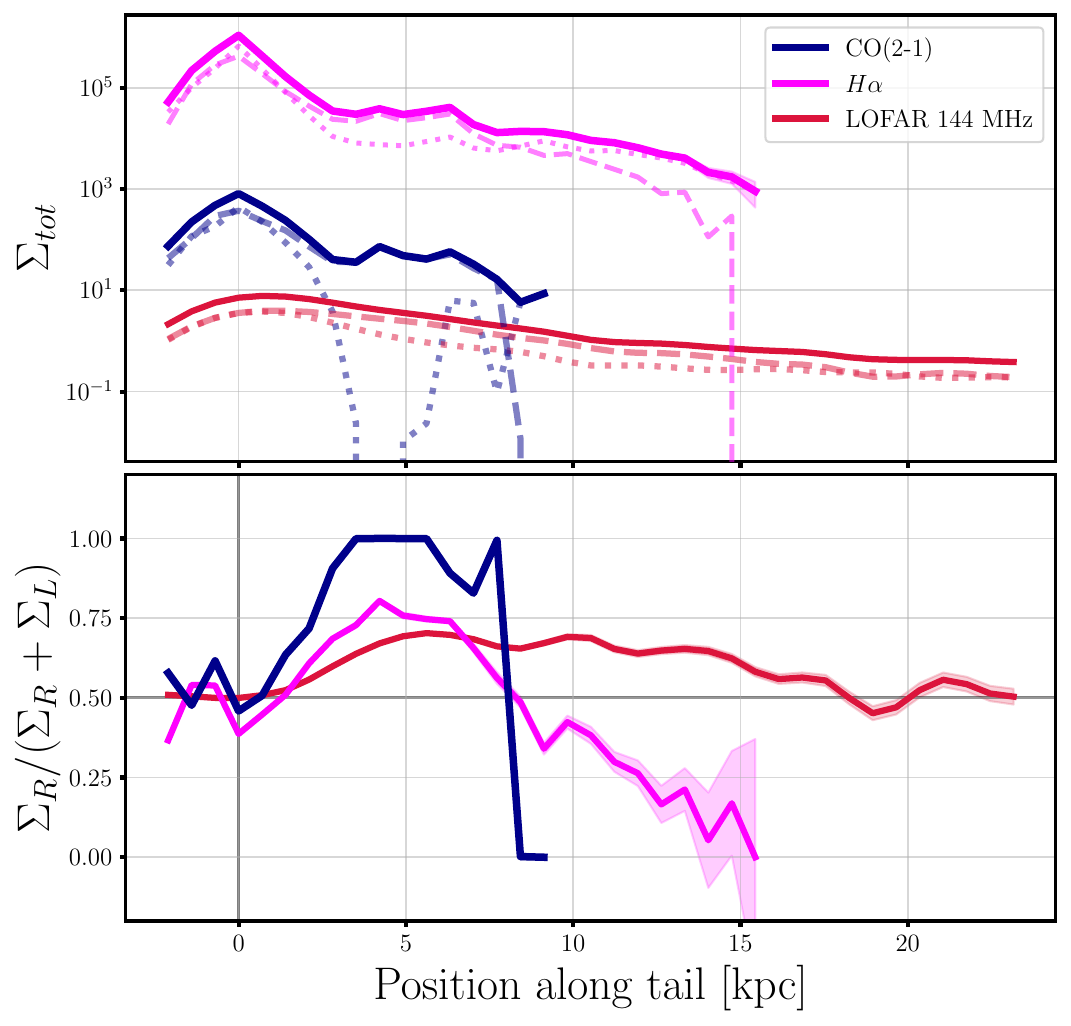}
    \includegraphics[width=0.39\textwidth]{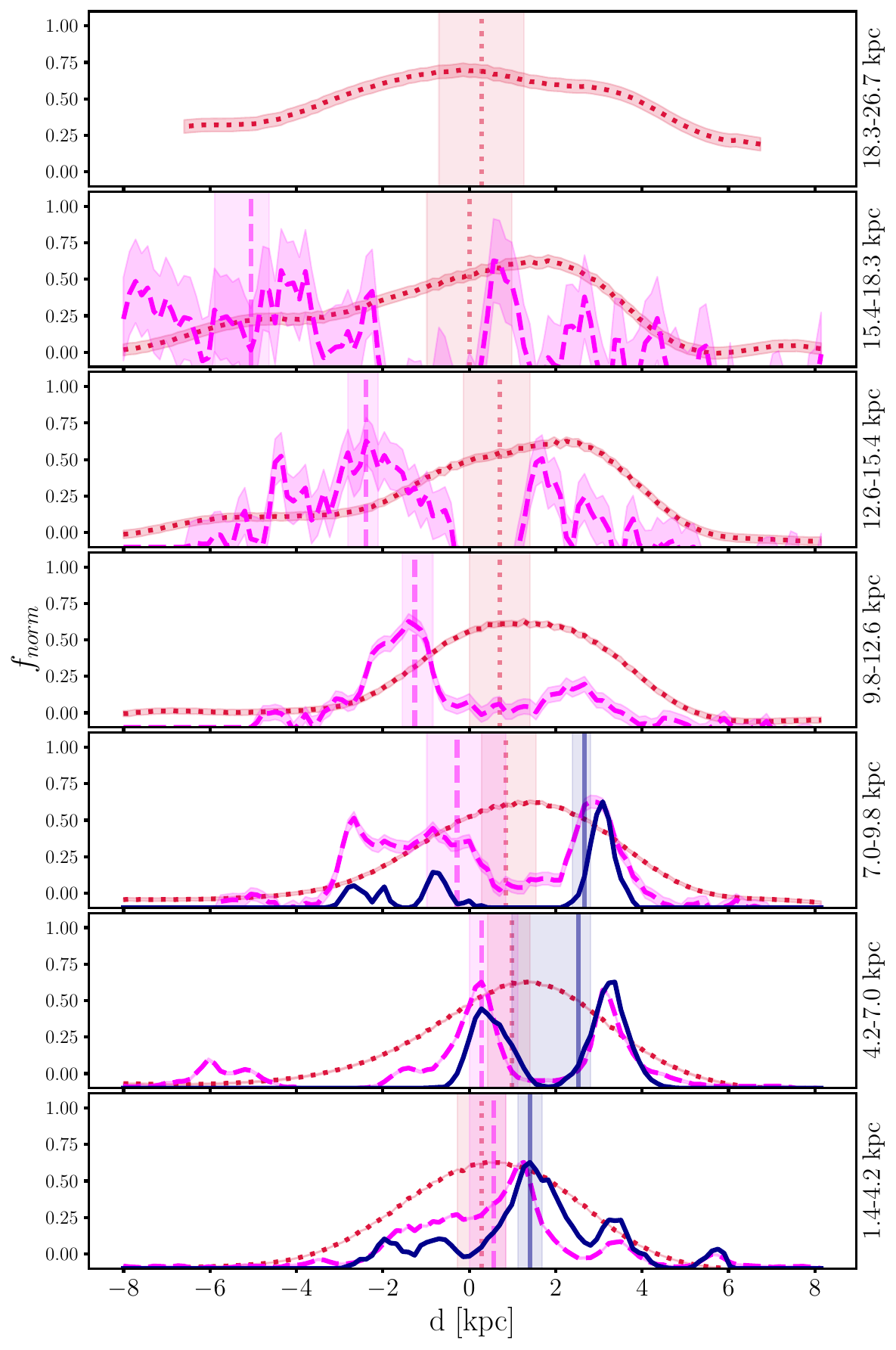}
    
    \caption{Asymmetry in tail flux distributions for CO (dark blue), \halpha{} (magenta) and LOFAR 144 MHz (red). \textbf{Left}: The total summed flux (top) and the asymmetry measurement (bottom) for each tracer, measured by the proportion of flux to the right (west) of the wind vector.     
    \textbf{Right:} The proportion of flux across slices perpendicular to the tail direction at varying distances along the tail, moving east to west across each slice. The flux-weighted centroids for each slice (and $\Bar{f} \pm10\%$ flux regions) are indicated as vertical lines. The topmost panel shows the summed flux for the remaining apertures for the LOFAR data.}
    \label{fig:tail-asymmetry}
\end{figure*}

We wish to measure asymmetries relative to the projected orbital path. We think that the outer tail is the best tracer of the orbital path, and that the inner tail is asymmetric due to rotational effects which are largest in the inner tail. For the tail angle we adopt the centroid of the LOFAR radio continuum tail, since that is the tail component measured furthest from the galaxy. Figure \ref{fig:tail-asymmetry} measures the flux in slices across the tail, increasing in distance along our estimated wind direction. We find a position angle of of $\theta_\text{wind} = -15\degree$ equally divides the radio continuum flux at distances 18-27 kpc, and adopt this as the tail angle. 

Furthermore, small dust filaments seen in the northwest region of the HST color image likely trace the local flow direction, which approximately aligns with the tail position angle measured from the radio continuum tail. There are some elongated dust and stellar features closer to the inner disk with different angles. However, such features do not always trace the large-scale wind angle, since the local flow direction close to the disk can be different due to various effects including rotation and shielding \citep{abramson_hubble_2014}.


\subsection{CO Brightness Azimuthal Distribution}
\label{sec:co-azimuth-dist}

The combined effects of ram pressure and rotation are expected to produce gas distributions that are more complex than simple head-tail morphologies which are predicted for face-on ram pressure interactions. In particular, large asymmetries may occur in the inner tail.

\begin{figure}
    \centering
    \includegraphics[width=0.45\textwidth]{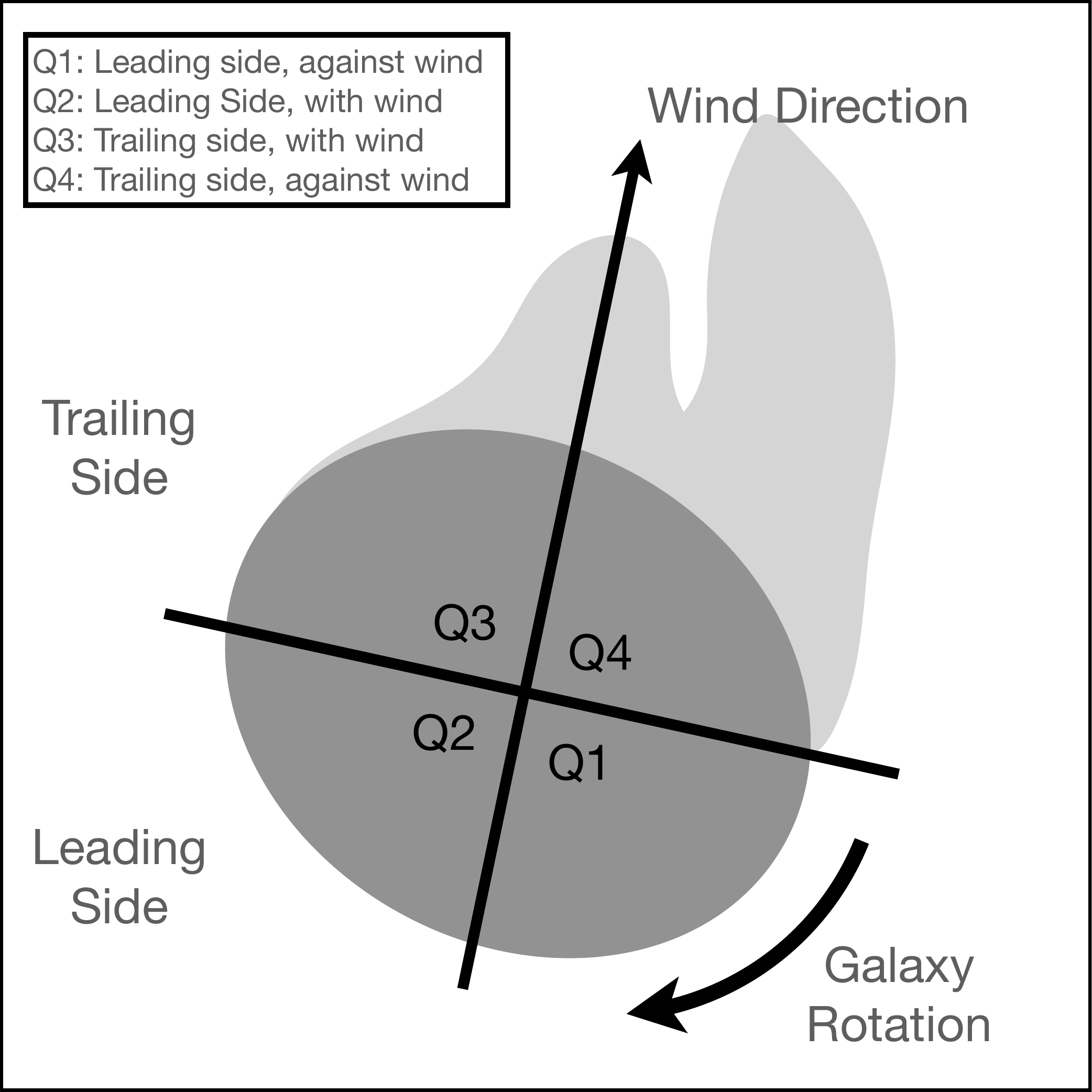}
    
    \caption{A schematic of the face-on view of the galaxy, with the quadrant divisions indicated. We note that the orientation of the quadrant boundaries depends on the estimated plane-of-sky wind direction, not the position angle of the galaxy's stellar disk. We also note whether each quadrant is rotating with or against the wind.}
    \label{fig:quadrants}
\end{figure}

To better understand the combined effects of ram pressure and rotation on NGC 4858, we analyze the CO spatial distribution in the galaxy outskirts as a function of azimuthal angle. We achieve this by placing elliptical annuli of increasing radii, and with the same position angle and ellipticity as the stellar disk, as measured in Section \ref{sec:stellar-distribution}. To explore the possible effects of compression, the outer radius of the innermost annulus is set to the radius of the truncated CO disk. We then take the median flux in regions of $\theta \pm 5\degree$. We estimate the uncertainty using the 2D uncertainty map generated by \textsc{Maskmoment} which combines the datacube's uncertainty level and the number of unmasked channels per pixel. We then multiply this by the square root of the area of each azimuthal slice.

 We bisect the flux distribution twice, first along the leading/trailing sides, and then along the approximate wind direction. This leaves us with 4 quadrants, which are shown in Figure \ref{fig:quadrants}. The quadrants are ordered to follow the galaxy's rotational direction, which is clockwise in the case of NGC 4858. Finally, the first two quadrants are the leading-side quadrants. The strength of local ram pressure varies between different quadrants, being strongest on the unshielded (leading) side rotating into the wind, and being weakest on the shielded (trailing) side rotating with the wind. 


\begin{figure*}
    \centering
    \includegraphics[width=\textwidth]{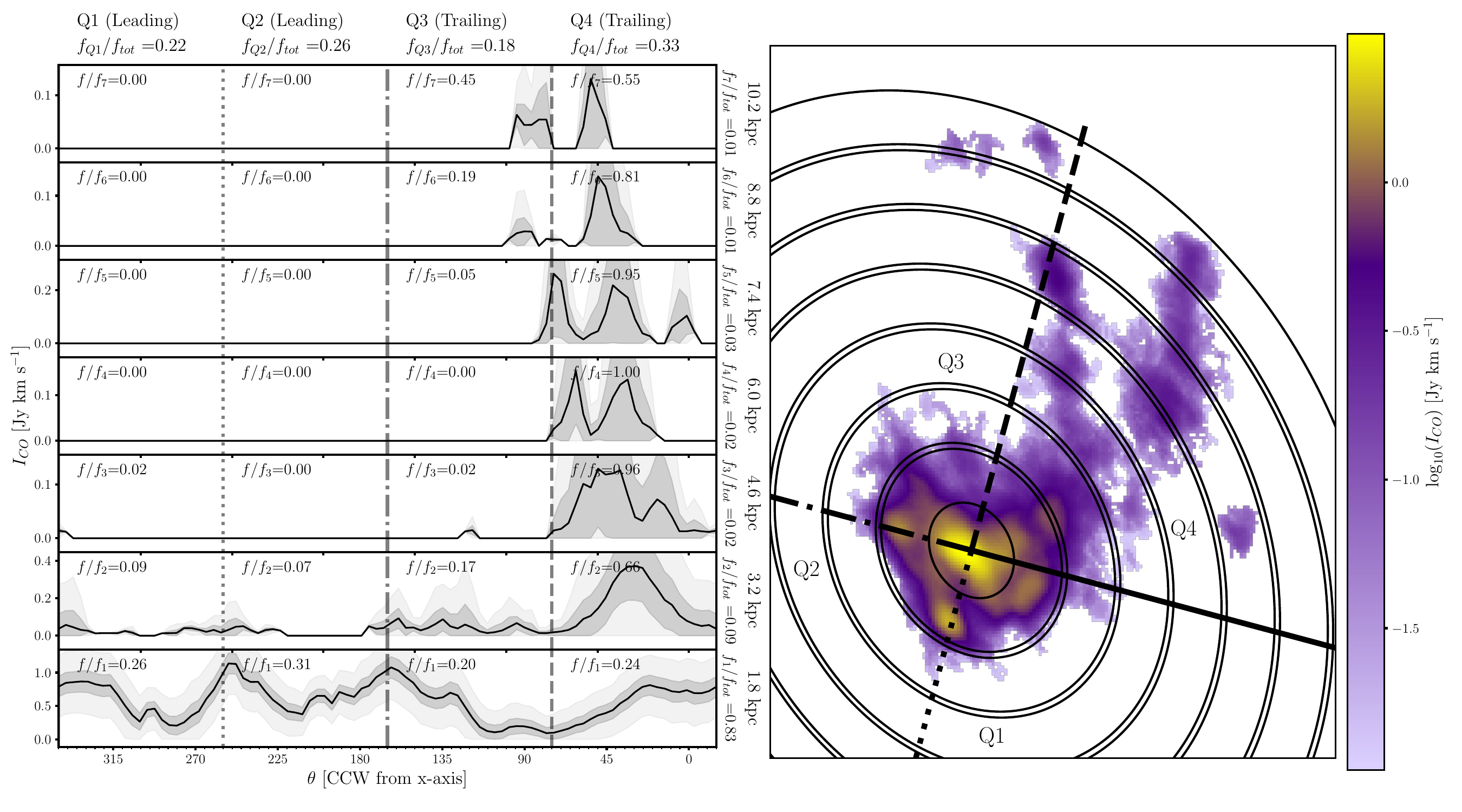}
    
    \caption{Azimuthal analysis of the CO flux distribution of NGC 4858. \textbf{Left}:  The flux and uncertainty contours  ($1\sigma$ and $3\sigma$) as a function of $\phi$ is shown for each annulus. Note that the vertical scale changes between annuli. The quadrant boundaries are denoted by the different line styles. The indicated radii correspond to the average of the inner and outer semi-major axes of each annulus.  The values $f / f_i$ in each box represents the division of flux across quadrants for each annulus, whereas $f_{Qi} / f_{tot}$ and $f_i / f_{tot}$ are the divisions of flux in each quadrant and annulus, relative to the total flux. \textbf{Right}: The elliptical annuli (overlaid on the CO moment-0 map) are oriented to correspond to the ellipticity and PA of the stellar disk, as determined by isophote fitting in Section \ref{sec:stellar-distribution}.}
    \label{fig:azimuthal-flux}
\end{figure*}

 Figure \ref{fig:azimuthal-flux} shows the results of this analysis, which offers a few key insights. The innermost annulus at $R=1.8$kpc shows more flux in quadrants 1 and 2 (the leading side) than quadrants 3 and 4 (the trailing side). The ratio of fluxes in the innermost annulus for these two sides is $F_\text{lead}/F_\text{trail} = 1.25$. Correspondingly, the three brightest CO peaks are near the truncated southern edge, and the brightest peak is near the southern leading point. 
 This agrees with the expected gas compression from a highly-inclined ram pressure event, where a large portion of gas found on the leading side will be pushed further into the disk, whereas gas on the trailing side has less impeding gas along the wind direction and can therefore be accelerated more efficiently.



Flux in the galaxy disk outskirts ($R=4.6$kpc and beyond) is almost entirely concentrated in quadrant 4, or the trailing-side quadrant rotating \textit{into} the wind. Indeed, almost $90\%$ of the CO flux in the outskirts is found in this quadrant. Quadrant 3 also has some flux in the innermost annulus, which is evidence of the expected elongation of gas along the trailing side overall, and also in the outermost annuli as some distant tail clouds are found there. 

Also seen is a trend in the azimuthal location of the tail peaks, which shift counterclockwise as a function of increasing radius. The gas tail also becomes more laterally symmetric at larger distances, changing from residing completely in Q4 near the base of the tail, to $\sim45\%$ of the CO flux residing in Q3 at the outermost annulus. Given that this galaxy has spiral arms, this is likely caused by a combination of inherent spiral structure and rotation, which is expected to be less important in the outer RPS tail. These peaks would not change location for a more uniform, straight gas tail. The peaks also appear to become more separated as the annulus radius increases.

\subsection{Discussion of Notable Features}
\label{sec:individual-feature-morphology}

In the following section we describe and discuss notable morphological features in the CO distribution. We also discuss how these features appear at different wavelengths, including Subaru \halpha{} and HST broadband optical.

\subsubsection{Southern Stripped and Compressed Leading Side}
\label{sec:morph-southern-leading-edge}

The CO flux contours along the southern side of the CO gas disk are considerably more bunched-up in this area than on the trailing side, suggesting that much of this gas is being pushed into the gas disk, compressing the gas. This is further evidenced by the stronger CO emission along the leading side of the galaxy (quadrants 1 and 2) in the $R=3.2$kpc annulus in Figure \ref{fig:azimuthal-flux}. There also appears to be strong \halpha{} in this region which might be caused by increased star formation caused by gas compression. Shock ionization of gas at the ISM-ICM interface is also a possible contributing factor, as this region of the galaxy is directly exposed to (and therefore strongly mixing with) the ICM wind. This region, however, also corresponds to a spiral arm in the galaxy, so it is hard to determine exactly how much of the amplified \halpha{} emission is due to ram pressure directly.

The southern half of the inner stellar disk ($r<3$kpc) shows a part of a spiral arm lacking both \halpha{} and CO in the region. This can be seen in Figure \ref{fig:overlay-maps}b,d.
This part of the arm has diffuse stellar emission, and is quite blue in color with $m_{350} - m_{600} = -0.75$. There are a few blue stellar point sources in the southern arm, but they are much less luminous than those in the northern part of the arm with strong CO emission and dust extinction. This indicates that star formation has largely stopped because of recent gas stripping, and not enough time has passed for the spiral structure to be dissipated via dynamical heating. It can also appear more blue due to dust having been stripped out of this region.

\subsubsection{Western Blobs}
\label{sec:morph-western-blobs}

To the west of the galaxy center are two \halpha{} clouds, both with head-tail morphologies, labelled B1 and B2 (See Figure \ref{fig:overlay-maps}c). The blobs are at $r=4.40$ and $r=6.13$ kpc, both beyond the CO gas truncation radius. These clouds are both located in Q4, the trailing side quadrant rotating into the wind. The location of these clouds suggests that they are not being shielded by the galaxy disk, and are therefore experiencing strong ram pressure. Interestingly, only the western feature shows a clear CO detection. B1, the eastern feature with no detected CO emission, is brighter (by 1.2 mag in 350LP) and bluer than B2 ($m_{600} - m_{350} = 0.49$ and $1.34$ for B1 and B2, respectively).
The different CO/\halpha{} ratios ($\sim6\times$ higher for B2) between these two blobs is suggestive of them being at different evolutionary stages. The cloud with both high amounts of CO and \halpha{} is likely at an earlier evolutionary stage, whereas the cloud with a lower CO/\halpha{} ratio is more advanced, having exhausted more of its available CO reservoir \citep[e.g.][]{poggianti_gasp_2019-1}.

\subsubsection{Tail Arms and Clouds}
\label{sec:morph-tail-structure}

The inner molecular gas tail of NGC 4858 has two distinct arm-like features, both approximately 6 kpc in projected length. Both tail components appear to originate from approximately the same location of the stellar disk, but this might be due to projection effects. Both tail components appear relatively straight in projection. The western tail is roughly parallel to the ram pressure wind vector, and the eastern tail is offset by approximately $20\degree$.

The tail is also prominent in \halpha, which is highly asymmetric in the inner $\sim4$ kpc. There are prominent \halpha{} counterparts to the 2 main CO features in the inner tail, but the relative \halpha{} and CO distributions in the 2 arms are strikingly different. The westernmost component shows good correspondence between CO and \halpha. The central component is quite broad in \halpha, with a width of approximately 3kpc. There appear to be 2 chains of bright features at either side of this feature, with fainter emission in between. There is CO detected only on the west side of the arm, or upstream assuming the tail features are rotating from east to west, from the \halpha{} peaks. It might be the case that these are two separate stripped components that have since been pushed together while rotating towards Q4.

North of these two prominent ``bunny ear" tail components ($R \sim 10$kpc) are 2-3 gas clouds, each approximately 1 kpc in diameter. They are mostly irregular in morphology, and fainter than the inner tail arms. The western cloud is brighter than its eastern counterpart, which might be two separate clouds that are overlapping with each other.
Based on their location, it is possible that these more distant clouds are related to the eastern tail arm, as they reside in the same line as that tail arm.

There is also a weaker eastern \halpha{} arm with no detected CO counterpart that becomes moderately strong 7-10 kpc from the galaxy center. The arm is formed from a chain of compact \halpha{} blobs which are likely HII regions, as they coincide with bright blue stellar complexes as seen in Figure \ref{fig:overlay-maps}b. There must be some molecular gas associated with these star-forming regions, but this is apparently below the ALMA detection threshold. 

While \halpha{} emission is more widespread in the tail than the detected CO emission, the \halpha{} emission in the inner tail (within 7kpc) is also highly asymmetric, and strongest in Q4 (See Figure \ref{fig:tail-asymmetry}).  The strong concentration of dense gas in Q4 in the inner tail is a topic we explore further in the following sections.

\section{CO Kinematics}
\label{sec:co-kinematics}

\begin{figure*}
    \centering
    \includegraphics[width=\textwidth]{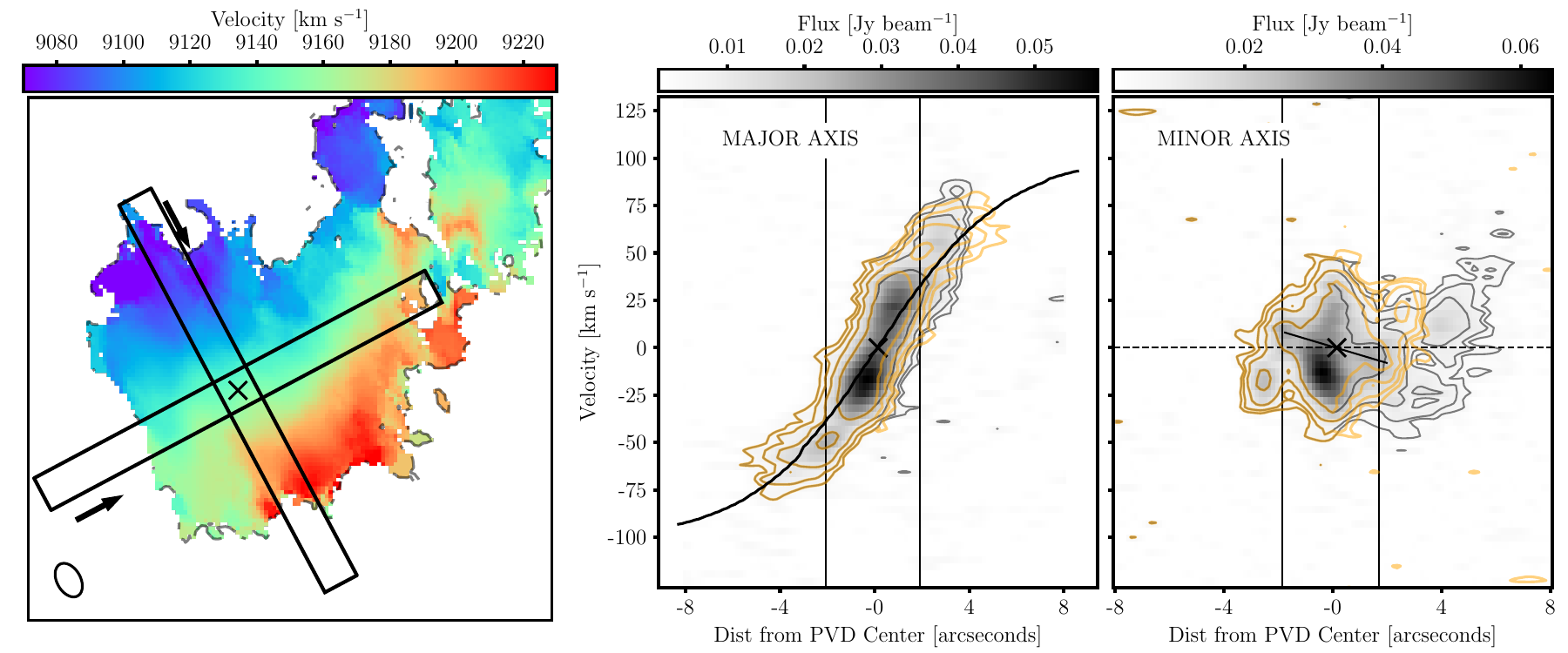}
    \caption{Position-Velocity Diagrams (PVDs) for the major and minor axes of NGC 4858. The PVDs are centered on the optical distribution, which we indicate in all three panels with an x. The arrows indicate the direction of each PVD slice (from left to right). The width of the PVD slice corresponds to the beam major axis ($\sim 1"$). 
    Orange contours are the contours for the leading side of the PVD, which are then flipped and mirrored to show offsets between the leading and trailing sides. 
    The contour levels correspond to [2, 3, 5, and 10] $\sigma$. The horizontal dashed line represents the systemic velocity of the galaxy. The estimated rotation curve (Section \ref{sec:residual-velocities}) is indicated in the major axis PVD with a solid black line. The vertical solid lines indicate the extent of the bar region. The velocity gradient that might be caused by bar streaming motions is also indicated with a solid line in the minor axis PVD.}
    \label{fig:pvds-majmin}
\end{figure*}

Having assessed the overall structure of NGC 4858 and its RPS tail, we now examine the kinematic features of the CO gas. Figure \ref{fig:momentmaps} shows the moment-1 velocity and moment 2 velocity dispersion maps for the galaxy. The moment 1 map shows characteristics of rotation throughout the galaxy's gas disk, with overall blueshifts in the northeast to redshifts in the southwest. 
The moment 2 map reveals a velocity dispersion of 15-30 \kms{} throughout the CO distribution except for the region where the tail arms "connect" to the CO disk in Q4, the edge of the CO disk in Q3, and along the leading side of the gas disk in Q1. Higher velocity dispersion in these areas may either be due to higher turbulent motion perhaps caused by mixing with hot ICM gas at the edges of the CO gas disk, or the presence of multiple, individually-coherent velocity components. 

Figure \ref{fig:pvds-majmin} show position velocity diagrams (PVDs) with slices across the major and minor axis, with slice widths equivalent to the beam major axis ($1.35" = 630$pc). The PVDs are centered on the stellar light peak, as measured using the Subaru R-band image. The extent of $3\sigma$ emission in the major axis PVD covers a velocity range of -70 \kms{} to 90 \kms with respect to the systemic velocity of $\sim9150$ \kms. The major axis PVD goes through the two brightest CO peaks, which are offset from the stellar light center. This is a feature commonly seen in barred galaxies \citep{kenney_twin_1992}.

The major and minor axis PVDs both show evidence for non-circular motions. Asymmetries can be seen across the major axis, where the blueshifted side of the PVD appears slightly more extended, and appears to flatten more than the redshifted side. In the minor axis PVD, the inner 2" shows a symmetric pattern of non-circular velocities with a velocity gradient across the nucleus with an amplitude of 8\kms. These are consistent with bar streaming motions, and are spatially coincident with the bar. The moment 1 map does not show the Z-shaped distortions in the velocity field characteristic of a bar \citep{sellwood_dynamics_1993}. This could be due to tail gas in front of the disk complicating the velocity field, as suggested by the complex dust morphology near the the galaxy center, or possibly disturbances to the gas kinematics near the galaxy center.

Beyond the central 2", the minor axis PVD shows non-circular motions likely associated with ram pressure. On the eastern, leading side, gas is blueshifted by $\sim20$\kms, in the expected direction for ram pressure acceleration. On the western, trailing side, the lines are broad with multiple velocity components. There is a blueshifted component: possibly gas accelerated by ram pressure. However, there is a strong redshifted component, corresponding to gas moving inwards (described later in this section). These components are likely at different distances along the line of sight. Possibly the blueshifted component is in the tail projected in front of the disk, and the redshifted component is associated with the disk and the western "bunny ear" arm. In the moment 1 map, these components are averaged together, but since the redshifted component is stronger there is a net redshift in the mean velocity.

\subsection{Residual Velocities}
\label{sec:residual-velocities}

\begin{figure*}
    \centering
    \includegraphics[width=\textwidth]{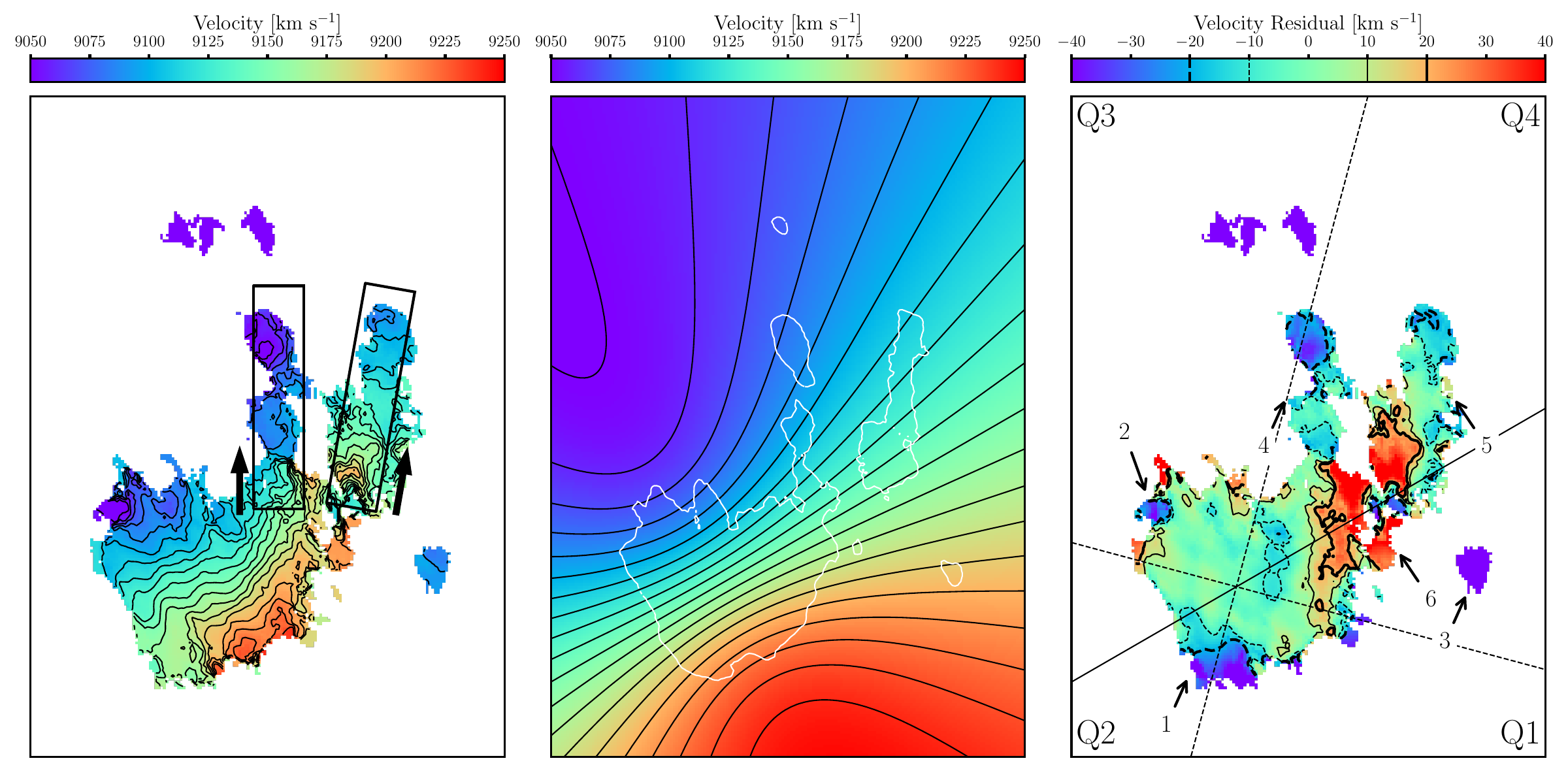}
    \caption{Velocity maps for NGC 4858. \textbf{Left}: The moment-1 velocity map with isovelocity contours at 10 \kms{} spacing. The PVD slices through the tail (See Figure \ref{fig:pvds-tail}) are shown as black boxes. \textbf{Center}: The circular velocity model with the same isovelocity contours, and a 0.1\jykmsbeam{} isophotal contour to show the CO footprint. \textbf{Right:} The velocity residual map with isovelocity contours at $\pm 10$ \kms{} to highlight deviations from circular motion. The quadrant divisions are shown as dashed lines, and the kinematic minor axis (where residuals are not affected by changes in the model rotation curve) is shown as a solid line. Annotations are included to highlight features with high velocity residuals that are explored in later sections. 
    1: Compressed leading side (Section \ref{sec:feature-leading-edge}). 
    2: Northern disk blueshifted complex (Section \ref{sec:feature-north-disk-blob}).  
    3: Western blueshifted blob (Section \ref{sec:feature-western-blobs}). 
    4 and 5: Main tail components (Section \ref{sec:feature-bunny-ears}). 
    6: Redshifted inner tail material (Section \ref{sec:feature-inner-tail-redshifted}).} 
    \label{fig:velocity-residual}
\end{figure*}

\begin{table}[]
    \centering
    \begin{tabular}{c|c c c}
         Residual & Location & Source & \\
         \hline
                    & Above Disk        &   Outward       & $\Hat{z}$ \\
                    & Below Disk        &   Inward       &  $\Hat{z}$ \\
         \textbf{Blueshift}   &  Major Axis (Approaching)  &  Faster       &  $\Hat{\theta}$\\
                    &  Major Axis (Receding)  &     Slower    &  $\Hat{\theta}$\\
                    &  Minor Axis (Near Side)  &  Outward       &  $\Hat{r}$\\
                    &  Minor Axis (Far Side)  &   Inward      &  $\Hat{r}$\\        
                    \hline
                    & Above Disk        &  Inward    & $\Hat{z}$    \\
                & Below Disk  & Outward  &  $\Hat{z}$         \\
            \textbf{Redshift}        &  Major Axis (Approaching)  &  Slower       & $\Hat{\theta}$ \\
                    &  Major Axis (Receding)  &  Faster       & $\Hat{\theta}$ \\
                    &  Minor Axis (Near Side)  & Inward        & $\Hat{r}$ \\
                    &  Minor Axis (Far Side)  &  Outward       &  $\Hat{r}$ \\

    \end{tabular}
    \caption{Possible sources for redshifted/blueshifted velocity residuals. For azimuthal velocities, "faster" and "slower" correspond to velocities faster/slower than the expected circular velocity given by the galaxy's rotation curve. See Appendix \ref{app:high-velocity-residuals} for a more detailed discussion.}
    \label{tab:velocity-residuals}
\end{table}

To best reveal non-circular motions which may be caused by RPS, we first subtract circular motions using a simple circular-velocity model and explore features revealed in the residual map. 
The location of the features relative to the major/minor axes of the galaxy provides information on the overall kinematics of these gas clumps.

Residual velocities may correspond to vertical, azimuthal, or radial motions depending on their location relative to the kinematic major or minor axis. Assuming that the motions lie within the disk plane (which is a reasonable approximation given a highly-inclined disk-wind angle which will keep gas near the disk plane), these residuals will correspond to motions outlined in Table \ref{tab:velocity-residuals}. More details are discussed in Appendix \ref{app:high-velocity-residuals}. 

To obtain a residual velocity map for NGC 4858, we choose to fit a simple circular velocity model to the moment-1 CO velocity map directly. We are concerned with how the CO gas velocities compare to the simplest model of the pre-stripped galaxy (i.e. only circular motions with a typical rotation curve). To obtain the rotation curve we use the \Barolo{} fitting package \citep{di_teodoro_3d_2015}. We use the ellipticity and position angle of the stellar disk obtained by isophotal analysis as our inputs (Section \ref{sec:stellar-distribution}). We fix these parameters, as well as $x_0$, $y_0$, and $V_\text{sys}$ and assume no radial velocities. This provides us with estimates for the rotation curve within the CO truncation radius, but we seek to estimate the rotation curve out to the outskirts of the stellar disk. To do this, we fit a simple universal rotation curve (URC) fitting function given by

\begin{equation}
    V(r) = V_\text{max}\tanh\left(\frac{r}{r_t}\right)
\end{equation}

\noindent where $V_\text{max}=165$\kms{} is the inclination-corrected asymptotic maximum velocity which we estimate using the \textit{i}-band galaxy luminosity \citep{alam_eleventh_2015} 
and the Tully-Fisher relation \citep{tully_new_1977}. The transition radius $r_t$ adjusts the scaling of the rotation curve, which we leave as a free parameter when fitting the model. Our best-fit to the available datapoints from \Barolo{} returns a transition radius of $r_t=2.4$kpc, similar to the CO truncation radius. From this, we are able to obtain a smooth rotation curve beyond the limitations of the \Barolo{} extracted datapoints. This is shown in Figure \ref{fig:pvds-majmin}. 

Since the rotation curve beyond the inner disk is somewhat uncertain, we have assessed the robustness of our derived velocity residuals by testing a range of outer galaxy rotation curves. In addition to our preferred rotation curve based on the Tully-Fisher relation, we also tried a rotation curve that is flat beyond the inner CO disk, and one with a maximum rotation velocity twice as high as expected from the Tully-Fisher relation. 
The corresponding velocity residual maps are shown in Appendix \ref{app:changec-velresids}.
None of the features we discuss with significant velocity residuals can be explained away by any plausible rotation curve, although the amplitude of the residuals does vary for some of the features. Additionally, features residing along the kinematic minor axis (such as the inner-tail redshifted feature) are insensitive to any changes in the rotation curve.



\begin{figure}
    \centering
    
    \includegraphics[width=0.5\textwidth, trim = 0mm 0mm 218mm 0mm, clip]{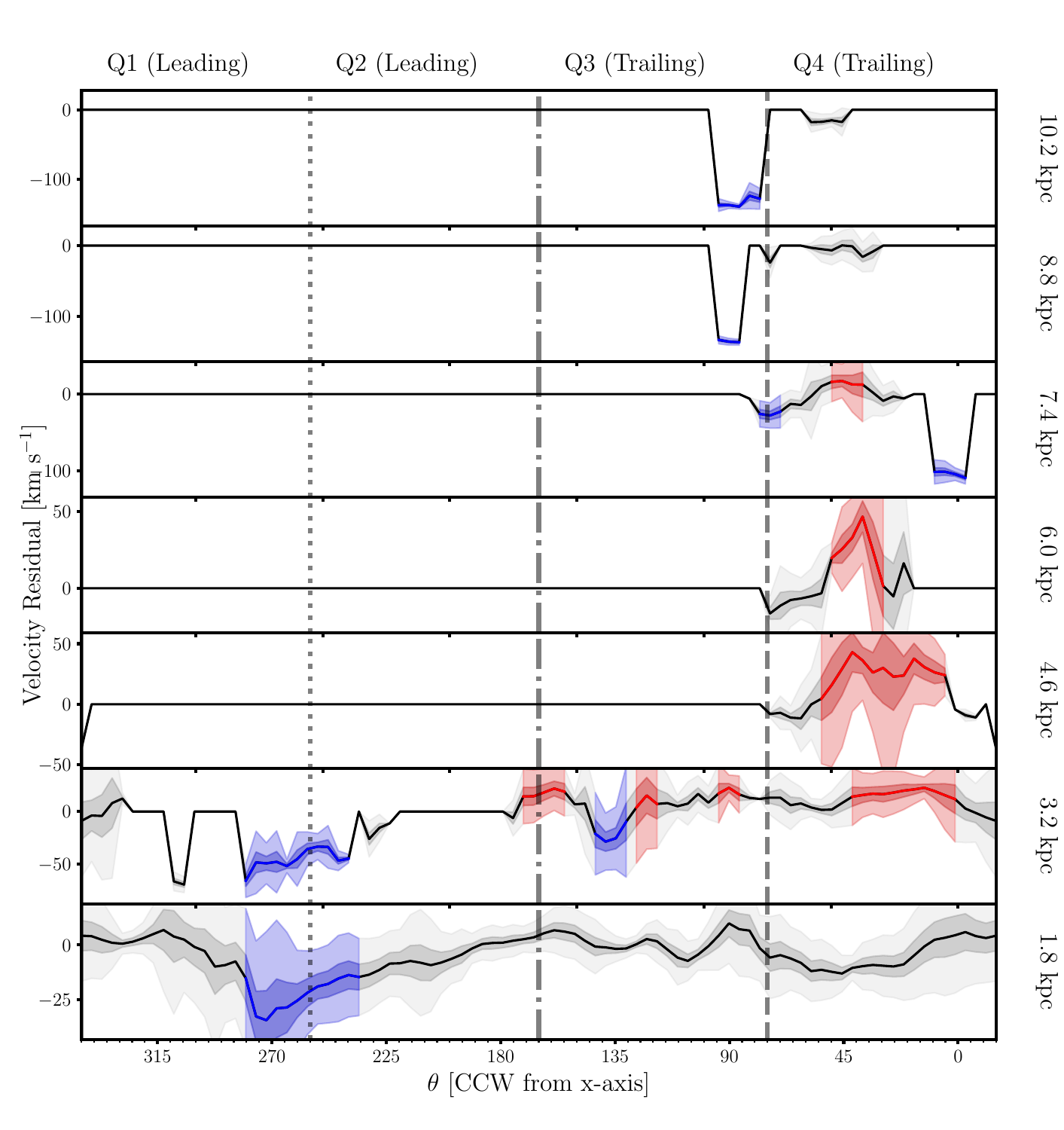}

    \caption{Azimuthal distribution of the residual velocity  map shown in Figure \ref{fig:velocity-residual}. The vertical scaling is different for each panel. The annuli and quadrants are the same as in Figure \ref{fig:azimuthal-flux}. Regions of higher velocity residuals ($|v_\text{res}| > 20$ \kms) are coloured blue or red to correspond to high residual blueshifts or redshifts, respectively.}
    \label{fig:azimuthal-velresid}
\end{figure}

We generate our velocity map using this rotation curve  with a simple disk rotation model that expresses line-of-sight velocities in polar coordinates ($r, \theta$) as 

\begin{equation}
    V_r(r, \theta) = V_\text{sys} + V_\text{circ}(r) \sin(i)\cos(\theta)
\end{equation}

\noindent where $V_\text{circ}(r)$ is the circular velocity, $i$ is the stellar disk inclination ($38\degree$), and $V_\text{sys}$ is the systematic velocity in \kms. The velocity model, as well as the residual map are shown in Figure \ref{fig:velocity-residual}. Residuals within the disk do not correspond with any patterns expected of errors in any of the model parameters \citep[See][]{van_der_kruit_kinematics_1978}. 
Most of the features with high residual velocities are blueshifted, as expected from ram pressure acceleration, but there is one large feature (located entirely in quadrant 4) with a highly redshifted residual velocity. Figure \ref{fig:azimuthal-velresid} shows the magnitudes of these residual features by repeating the azimuthal analysis described in Section \ref{sec:co-azimuth-dist} on the velocity residual map.


\section{Selected Features with High Velocity Residuals}
\label{sec:high-residuals}

Having assessed overall trends in velocity, we now turn to individual features with high velocity residuals. For each of these features, we determine what its residual velocities might correspond to in the galaxy's frame of reference. A more detailed explanation on what residual velocities might correspond to can be found in Appendix \ref{app:high-velocity-residuals}. We explore features with residual velocities higher than $\pm20$ \kms{} for analysis, and discuss the blueshifted features first.

\begin{figure}
    \centering
    \includegraphics[width=0.49\textwidth]{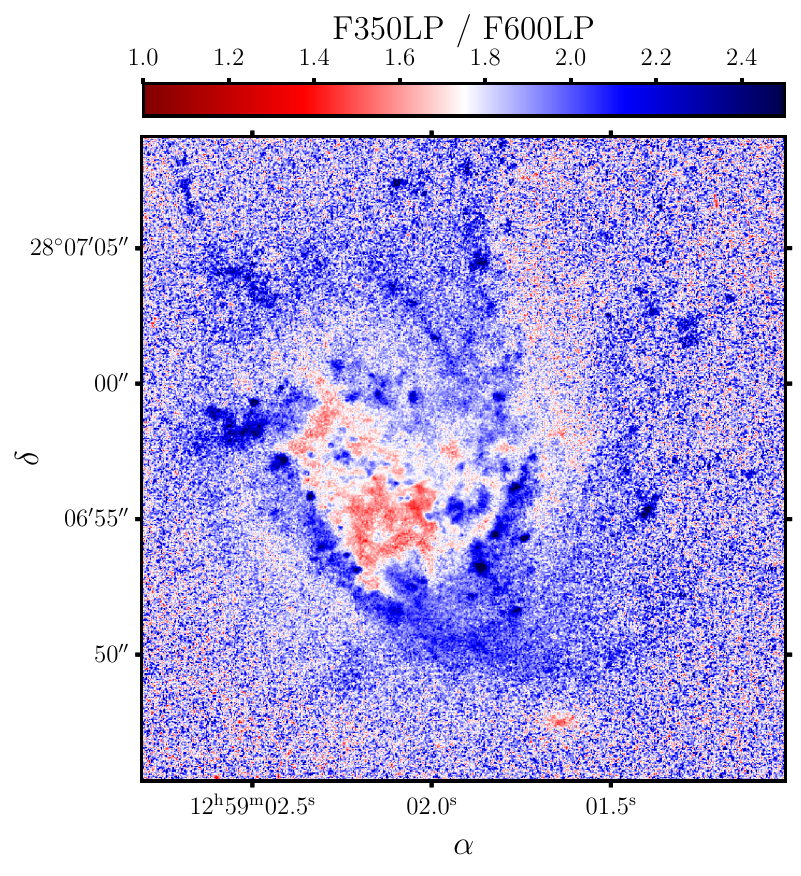}
    \caption{The ratio of the F350LP and F600LP HST images. The color scaling is adjusted to highlight dust reddening throughout the disk, as well as regions with bluer stars. }
    \label{fig:hst-color-ratio}
\end{figure}

We assume for all of these features that the tail is in front of the disk. Figure \ref{fig:hst-color-ratio} shows dust reddening  throughout the disk which coincides with the dust extinction seen in the HST color image. Furthermore, the tail lying behind the disk would require a total velocity through the cluster of $> 5000$\kms, more than $5\times$ the cluster velocity dispersion \citep{sohn_velocity_2017}. 

\subsection{Southern Leading Side}
\label{sec:feature-leading-edge}

The southern side of the CO gas disk close to the truncation radius of $\sim3$kpc and between $210-280\degree$ is blueshifted to between -25 and -50 \kms (ID 1 in Figure \ref{fig:velocity-residual}). This region extends approximately 1kpc radially into the gas disk. Based on its location and the ram pressure wind vector, it is easy to conclude that this is gas being accelerated directly by ram pressure, in a region of the galaxy not well-shielded from the wind. The highest level of blue-shifting in this region, at 40-50 \kms, is towards the southernmost edge of the disk.

\begin{figure}
    \centering
    \includegraphics[width=0.50\textwidth]{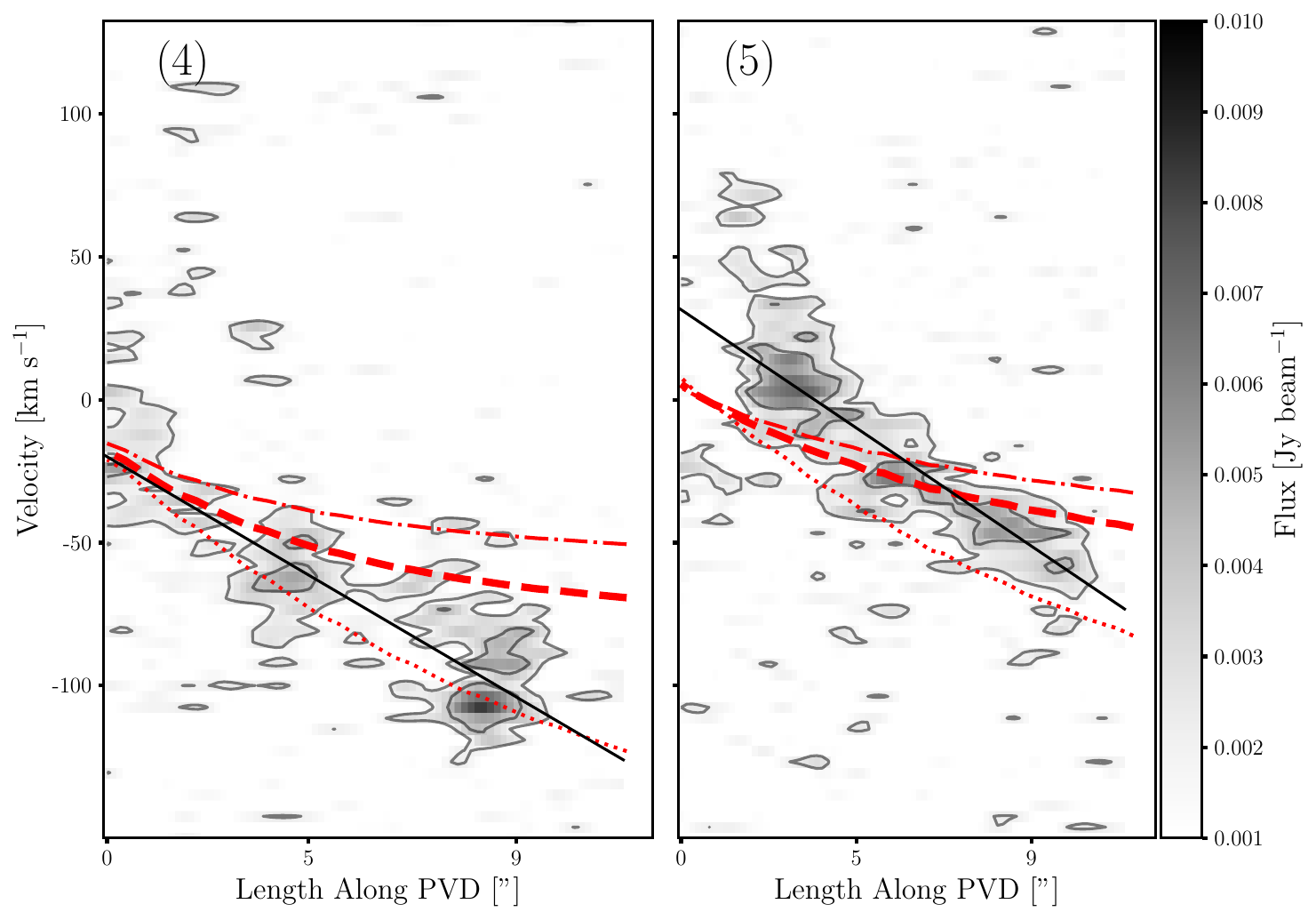}
    \caption{PVDs through the "bunny ear" tail components (See Figure \ref{fig:velocity-residual} for the positions of the slices). The PVD widths for both slices are twice the beam width. Both PVDs start closer to the disk and extend northward. The contour levels are the same as in Figure \ref{fig:pvds-majmin}. The solid black lines on the PVD indicate the fit to the velocity gradient along each tail component. The dark dashed red line indicates the expected circular velocity along each PVD, whereas the dotted and dot-dash lines are for a curve that immediately flattens beyond the truncation radius, and a curve with twice the $V_\text{max}$ expected from the Tully-Fisher relation, respectively.}
    \label{fig:pvds-tail}
\end{figure}

\subsection{Northern Disk Blobs}
\label{sec:feature-north-disk-blob}

On the northern side of the disk is a cohesive strongly blueshifted feature (ID 2 in Figure \ref{fig:velocity-residual}). This appears in the azimuthal analysis at the 3.2 kpc annulus, reaching a residual velocity of approximately -40 \kms. Assuming that any stripped material is on the near side of the disk, this might correspond to vertical motions moving away from the disk mid-plane. It does not reside on either axis, so both azimuthal (faster than rotation velocity) and radial (outward) motions can contribute. Regardless of a what type is contributing, this would correspond to gas clouds being stripped away from the gas disk.

\subsection{Western Blobs}
\label{sec:feature-western-blobs}

The one western blob with CO (B2, discussed in Section \ref{sec:morph-western-blobs}), has a very strong velocity residual of approximately -100 \kms. Based on its location, the residual velocity can be a combination of all three possible motions. Assuming this cloud is in front of the disk, this may correspond to outward radial motions, azimuthal motions slower than the expected circular velocity, or vertical motions moving away from the disk mid-plane. These motions would be that expected from a dense cloud experiencing strong ram pressure acceleration.

\subsection{Tail Components}
\label{sec:feature-bunny-ears}

Both tail components (IDs 4 and 5 in Figure \ref{fig:velocity-residual}) have an approximately linear velocity gradient. Figure \ref{fig:pvds-tail} compares the PVDs to the velocity gradient expected from rotation. The velocity gradient in both tail components is approximately -20\kms{} kpc$^{-1}$ with greater blueshifts away from the disk. These gradients are about 2 times larger than that predicted from rotation alone. The region around the bases of both arms has a larger velocity dispersion compared to the gas disk, which suggests either high turbulence or multiple velocity components. 

The eastern arm has velocities at its base consistent with rotation, whereas the outer arm has a gradually more blueshifted residual moving away from the disk. 
This is consistent with a relatively undisturbed base, but ram pressure acceleration of the outer arm, suggesting that the arm is being stretched apart by ram pressure. 
Towards the end of the tail component there appear to be two distinct clouds, contributing to higher velocity dispersion in this area.

The western arm has a similar velocity gradient,
but has an overall redshift, relative to both the eastern arm and the expected rotation velocities.
The outer arm has a modest blueshifted residual velocity,
but the base of the tail, which is near the minor axis, has redshifted residual velocities which we discuss in greater detail in the following section.
The western arm leads the eastern arm, and was previously at the location of the eastern arm.
This might suggest a similar history to the gas in the eastern arm, in which ram pressure acceleration of the outer arm created a larger velocity gradient, with higher blueshifts for the outer arm.
But now as this arm has rotated into the quadrant rotating into the wind, the entire arm is falling back toward the galaxy, giving it an overall redshift.

\subsection{Inner Tail Redshifted Gas}
\label{sec:feature-inner-tail-redshifted}

Contrary to the other features with high velocity residuals in NGC 4858, there is one major component that instead has \textit{redshifted} residuals (ID 6 in Figure \ref{fig:velocity-residual}). This feature is located around the base of the "bunny-ears" where the tail bifurcates, and extends down the western side of the disk. The residual velocities of this region range from $\sim20$ \kms{}, to upwards of $50-60$ \kms{} closer to the bifurcation.

Most of the redshifted material lies close to the minor axis. Since dust extinction implies that the tail is in front of the stellar disk, the redshifted gas would correspond to vertical motions moving towards the disk plane, or inward radial motions.  In the intermediate regions where redshifted gas is seen between the major and minor axis, this might also correspond to gas moving faster than the circular speed. Given that off-axis regions are connected to the regions on the minor axis, it is reasonable to assume that the minor-axis motions (vertically and radially inwards) are the primary motions throughout. Though we cannot determine whether vertical or radial motions are more important, these motions all correspond to material falling into the disk potential.


\section{Asymmetries Due to Highly-Inclined Ram Pressure}
\label{sec:asymmetries}

\begin{figure}
    \centering
    \includegraphics[width=0.49\textwidth]{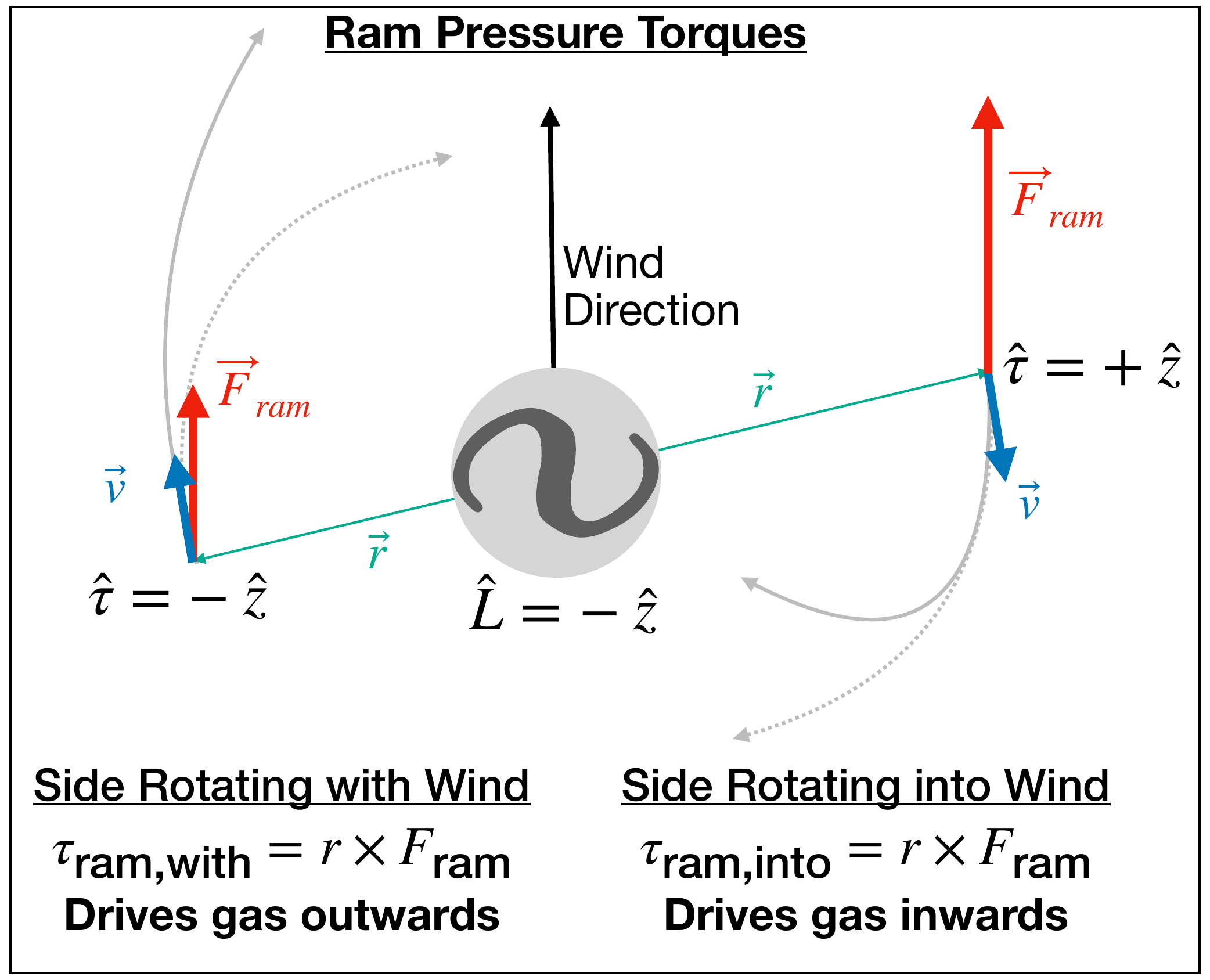}
    \caption{A diagram outlining wind-driven torques in the case of an edge-on wind for particles on opposite sides of the galaxy, looking from above along the z-axis. The dashed grey arrows show unperturbed orbits (dashed) orbits. The solid arrows are schematics indicating possible perturbed trajectories.}
    \label{fig:torques}
\end{figure}

The most striking characteristic of NGC 4858's inner RPS tail is its asymmetrical structure, extending from the western side of the disk. The formation of large asymmetries in the inner tail can be understood by analyzing the interplay of rotation and ram pressure on gas parcels orbiting around the galaxy disk. In this section we first estimate the 3D tail geometry of NGC 4858, and then describe the mechanism of RP-induced torques and their role in creating asymmetries in the CO gas distribution.

\subsection{Estimating the 3D Tail Geometry}
\label{sec:tail-geometry}

The disk-wind angle is the angle between the wind vector and the normal vector to the disk plane. The details of this calculation can be found in Appendix \ref{app:disk-wind-angle}.

The galaxy's gas tail has a plane-of-sky component extending northwards from the disk, as well as dust extinction and reddening that suggests the tail extends in front of the stellar disk (see Figure \ref{fig:hst-color-ratio}). This sets constraints on allowable tail angles and therefore total velocity of the galaxy. Under these simple constraints, the maximum range of disk-wind angles is  $38\degree$ (largest allowed face-on component) to $90\degree$ (completely edge-on).

To get an estimate for the total velocity of NGC 4858, we can use the galaxy dynamics package \textsc{Gala} to model the mass distribution of the Coma cluster \citep{price-whelan_adrngala_2024}. We model the Coma cluster mass distribution as an NFW profile with a total mass of $1.2\times10^{15}M_\odot$, with a concentration ($c=R_\text{vir}/R_s =9$) \citep{lokas_dark_2003}. Assuming that the projected distance from the cluster center is the true distance to the cluster center, the expected total velocity at NGC 4858's current location is 3410 \kms. We can also take the uncertainty on this estimate to be equivalent to the velocity dispersion of Coma. This corresponds to a velocity along the plane of the sky at $v_\text{POS} = 2402_{-1842}^{+1196}$ \kms, and a disk-wind angle of \DiskWindAngle{} degrees, or highly inclined. The uncertainty on this estimate is large, due to the many assumptions required of the galaxy's total velocity. 
However, the generous uncertainty estimate based on galaxy velocity dispersion should cover all such uncertainties.




\subsection{Role of Ram Pressure Torques}
\label{sec:torques}

While RPS is typically described using the Gunn and Gott criterion \citep{gunn_infall_1972}, this treatment is insufficient to fully describe how material is accelerated, and the subsequent observed gas distribution. Ram pressure affects the linear momentum of a gas cloud principally via mixing with ICM gas, where the velocity of a gas cloud can be described as the fraction of ICM gas present in the cloud \citep{tonnesen_its_2021}. However, despite ram pressure being weaker on the side rotating with the wind, it is evident from simulations that this is the region where ram pressure is more efficient at stripping material \citep{roediger_ram_2006, jachym_ram_2009}. The effects on angular momentum due to torques imposed on gas by the ram pressure wind need to also be considered.

Figure \ref{fig:torques} visualizes the different effects on two gas clouds located on opposite sides of the galaxy: one rotating into the wind, and one rotating against the wind. The strength of these torques will not be equal, as the local ram pressure acceleration on the side rotating into the wind is stronger than for the side rotating with the wind. For the cloud rotating with the wind, the torque vector aligns with the angular momentum vector, increasing angular momentum. This process helps to accelerate the gas into divergent, more extended orbits. This is an important mechanism for helping to push out gas in highly inclined stripping events where rotation cannot be neglected. This will be more efficient in regions where this torque is highest, where there is less gas downstream from the clouds, and where the gas is already moving in a similar direction to the wind (as that gas will require smaller changes in momentum in order to be stripped).

For gas rotating into the wind, the torque vector is opposite to the angular momentum vector, reducing the angular momentum and decreasing the kinetic energy of the gas particles. This promotes a torque that drives gas deeper into the gravitational potential \citep{akerman_how_2023}. This may be an important process for fueling AGN, which might help explain enhanced AGN activity observed in RPS galaxies \citep{radovich_gasp_2019}.



The importance of ram pressure torque-induced changes in angular momentum can be evaluated by considering the differences in local ram pressure on either side of the galaxy disk. Analytically, ram pressure is stronger on the side of the galaxy rotating into the wind. The strength of asymmetries in ram pressure strengths is given by 

\begin{equation}
    \mathcal{R} = 1 -  \frac{(V_{\text{ICM}} - v)^2}{ (V_{\text{ICM}} + v)^2}
    \label{eq:differential-ram-pressure}
\end{equation} 

\noindent where $V_{\text{ICM}}$ is the ICM velocity and $v$ is the local gas velocity. In the case of no complicating factors, such as a bow shock, the ICM velocity will be equal to the galaxy's velocity (i.e. $V_{\text{ICM},} = -V_{\text{gal}}$). In the case for NGC 4858 with a maximum rotation speed of 160 \kms{} and a minimum velocity through the cluster of 2420 \kms, this corresponds to a differential RP strength of $\mathcal{R} = 0.23$. 
If the strength of the ram pressure wind is reduced by a bow shock, self-shielding, or some other mechanism, the local ICM velocity $V_{\text{ICM}, i}$ is reduced, which then increases $\mathcal{R}$. The asymmetries in an inner tail are likely elevated to the effects of differential torques.


\subsection{Rotation-Induced Stellar Complexes}
\label{sec:rot-induced-complexes}

\begin{figure}
    \centering
    \includegraphics[width=0.49\textwidth]{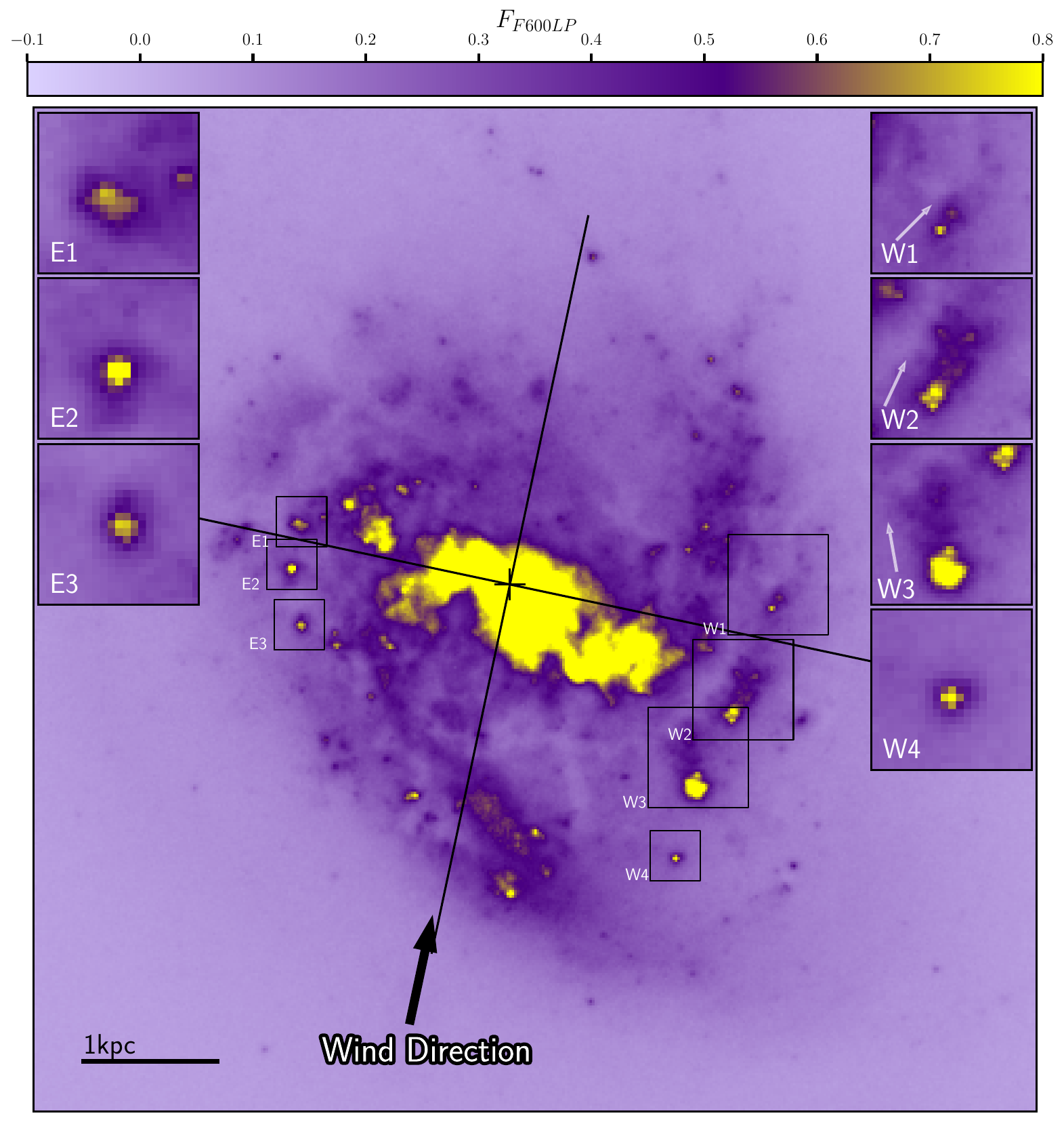}
    \caption{HST \textit{F600LP} image of NGC4858, divided into quadrants. The intersection of the quadrants is the stellar light centre from the Suprime-Cam R-band image, as found by isophote fitting and labeled by the cross. The approximate wind direction (NNE) is denoted by the arrow. 
    Regions of interest from each quadrant are shown with enlarged cutouts.}
    \label{fig:stellar-complexes}
\end{figure}

Figure \ref{fig:stellar-complexes} identifies in the HST F600LP image a collection of stellar complexes, coincident but close to the edge of the CO emission in the disk, with considerably different morphologies. Three of the complexes on the side rotating against the wind (W1-W3) have distinct head-tail morphologies. We do not identify any distinct features in the CO disk that relate to the stellar complexes, but this may be due to resolution and/or sensitivity limits in the ALMA data. All of the tails point northwards ($\theta_\text{tail} = -45\degree, -25\degree$ and $10\degree$ respectively). These structures appear to be extended along the same direction as the ram pressure wind. The tails have lengths of approximately 0.5 kpc. While W1 and W3 have mostly diffuse tails, W2 exhibits signs of clumpy structure within the tail, with multiple bright knots present. One other identified complex on this side of the galaxy, W4, does not show strong signs of a tail, but may be slightly extended towards the west. On the side rotating with the wind, complexes are also identified (E1-E3) but lack any clear tail structure. 

It may be initially surprising that these differences in morphology appear in \textit{stellar distributions}. The complexes can be seen as "mini jellyfish galaxies", where the same processes that create stellar tails in RPS galaxies are also occurring on a smaller scale in gas clouds near the stripping radius. On the side of the galaxy rotating into the wind, the local ram pressure is strong enough to ablate gas from the clouds. On the side rotating with the wind, the local ram pressure is sufficiently reduced by rotation to not create such an asymmetrical structure in accelerated gas clouds. The difference in local ram pressure, described in Equation \ref{eq:differential-ram-pressure}, appears sufficient in the case of NGC 4858 to induce such morphological differences.

\subsection{Evolution in the Inner Tail}
\label{sec:arm-formation}

\begin{figure*}
    \centering
    \includegraphics[width=\textwidth]{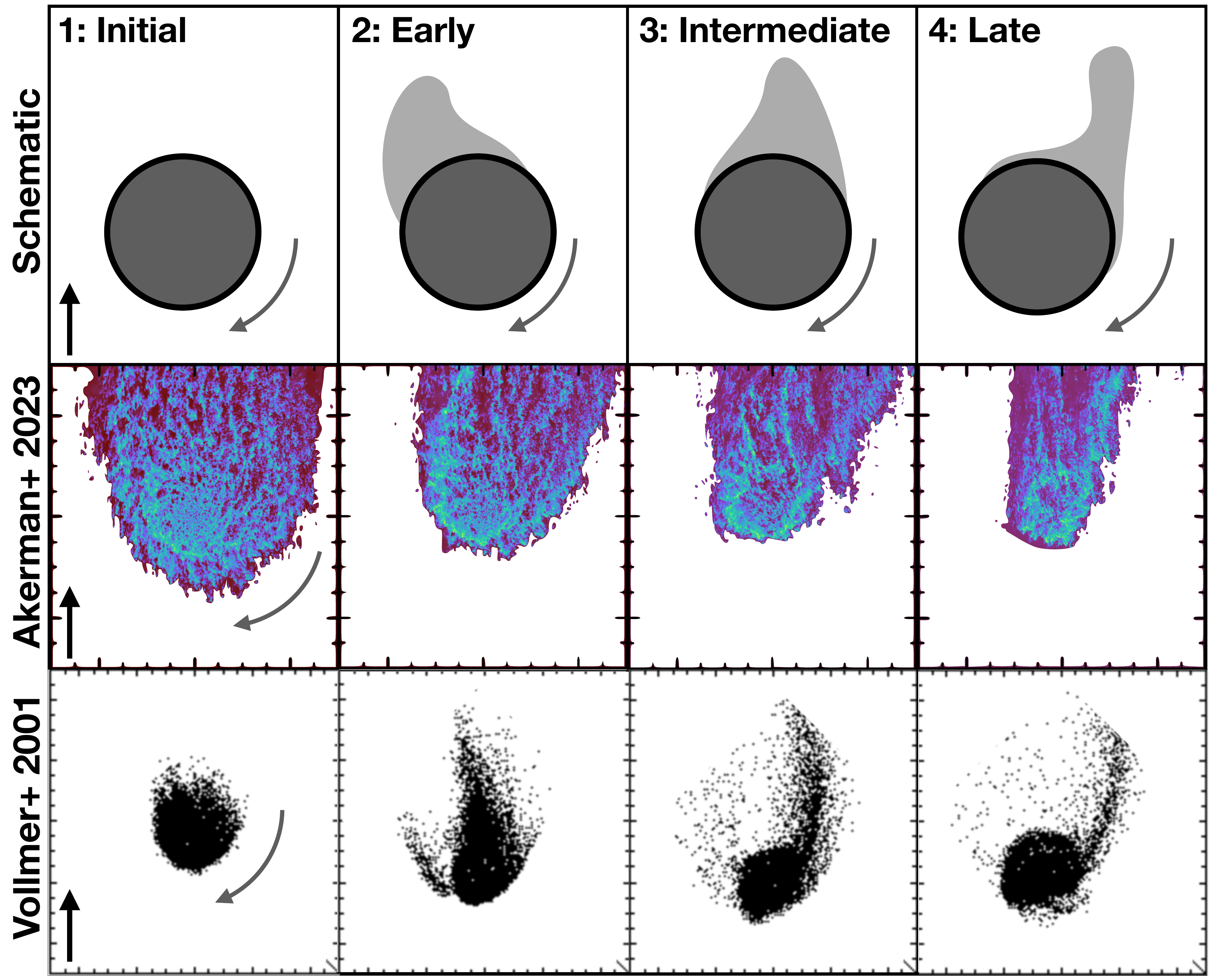}
    \caption{The development of an asymmetric tail due to a highly-inclined ram pressure interaction. The top row shows a general schematic of the process and the bottom two rows show this evolution in gas density in 2 different simulation types (Middle: Fig 1 of \cite{akerman_how_2023}, showing a hydrodynamical simulation and Bottom: Fig 11 of \cite{vollmer_ram_2001}, showing a "sticky n-body" simulation). The simulation snapshots have been rotated and flipped so that all galaxies are rotating clockwise, and the wind direction is pointed directly North.}
    \label{fig:arm-formation}
\end{figure*}

While ram pressure can accelerate and ultimately strip gas, it does not always accelerate gas to the velocity required to escape from the galaxy's gravitational potential well. In a scenario where gas is stripped from the disk but not accelerated to the escape velocity, this material has instead been accelerated to a highly elongated orbit, and may fall back into the disk.

Sticky N-body and hydrodynamical simulations show the development of an asymmetrical tail in the snapshots of a galaxy's evolution \citep{vollmer_ram_2001, lee_dual_2020,akerman_how_2023}. This confirms that an asymmetrical inner tail does not require hydrodynamical effects to develop. Fallback has been predicted by many simulations \citep{vollmer_ram_2001, tonnesen_gas_2009, akerman_how_2023, zhu_when_2023, sparre_magnetised_2023}, and has been observed in molecular gas observations of other RPS galaxies \citep[e.g.][]{cramer_molecular_2021}.

Figure \ref{fig:arm-formation} presents a schematic for the overall phases of inner-tail gas evolution and compares these phases to snapshots of two galaxy-scale simulations (one hydrodynamical, one sticky N-body) under highly-inclined ram pressure. Both of these simulations show the development of an asymmetric tail on a timescale of approximately 400-500 Myr, which is likely a similar timescale for NGC 4858's ongoing ram pressure event. The phases can be described broadly as:

\begin{enumerate}
    \item \textit{Early Phase}: Ram pressure initially blows out gas preferentially on the side of the galaxy rotating with the wind.
    \item \textit{Intermediate Phase}: Angular momentum moves the material from the side rotating with the wind, to the side rotating into the wind. 
    \item \textit{Late Phase}: The tail persists and concentrates on the side rotating against the wind. It is in this region where fallback is most likely to occur. 
\end{enumerate}

In the case of NGC 4858 and the two "bunny ear" tail components, it is likely that observations are showing pre-existing spiral arms that were stripped at different times and are now at different phases of inner-tail evolution. 
Comparing their locations with the phases in Figure \ref{fig:arm-formation}, it appears that the central tail component (ID 4) is in the intermediate phase and the other component (ID 5) is in the late phase. 
This implies that as evolution in this galaxy proceeds, the easternmost component might evolve to have similar morphological and kinematic characteristics to its counterpart.

While simulations tend to develop an arm in Q4, it should be noted that these simulations create such features with no pre-existing spiral structure in the initial, un-stripped galaxy disk. NGC 4858 on the other hand clearly has prominent spiral structure, so its evolution deviates somewhat from the simulation predictions. The same forces that act in Q4 to develop an inner tail arm will also act to strengthen any pre-existing spiral arm structure accelerated into the inner tail region.

\subsubsection{Later Evolutionary Signatures in NGC 4921}

\cite{cramer_molecular_2021} revealed the presence of fallback gas using ALMA CO(2-1) observations of the Coma galaxy NGC 4921. The fallback gas appears as three clouds on the leading side of the galaxy arranged in a spiral pattern, with an associated spiral arm segment of bright stars. These clouds are behind the stellar disk since there is no associated dust extinction. The blueshifted velocity residuals of the clouds, which are opposite in direction to residuals expected from ram pressure acceleration, are hypothesized to originate from vertical motions as the clouds fall back towards the galaxy disk mid-plane. Adopting our quadrant analysis, these clouds reside in Q1, but close to the Q1-Q2 boundary. Our proposed track suggests that these clouds might have fallen all the way back to Q1, having orbited across the inner tail. Because NGC 4921 is more massive ($M_B = -22.5$ compared to $M_B = -19.5$ for NGC 4858) there is a greater likelihood for dense clouds to persist all the way to Q1, as clouds will not be fully stripped as efficiently. It is likely that we are observing the same evolutionary process in both galaxies, but in the case of NGC 4921, we are witnessing even later evolution than what is captured in NGC 4858.


\section{Summary and Conclusion}
\label{sec:conclusion}


We have presented new high-resolution ($\sim1"$, $\sim500$pc) ALMA CO(2-1) observations of NGC 4858, a spiral galaxy in the Coma cluster. The galaxy is experiencing strong (and likely increasing) ram pressure at a large disk-wind angle, so the interaction between rotation and the ram pressure wind is strong. It is observed at an evolutionary stage where a large fraction of the galaxy's ISM has been recently pushed out of the disk, and where there is a large quantity of gas in the inner tail. Moreover, the galaxy is viewed relatively face-on, so we are able to study the interaction between rotation and ram pressure wind by analyzing the azimuthal gas distribution and kinematics. Using the CO observations alongside complimentary data, we analyze the evidence for a highly inclined, extreme RP interaction:

\begin{enumerate}
    \item \textbf{Asymmetrical Inner Tail:} Splitting the galaxy into 4 quadrants, aligned with the wind direction and trailing/leading sides, reveals that almost all of the CO tail gas is located on the trailing half of the galaxy, on the side rotating into the wind. Asymmetries in the inner tail are high for all measured gas tracers.
    \item \textbf{Velocity Residuals:} By subtracting a simple circular velocity model from the CO moment 1 velocity map, we find a wealth of features with non-circular motions around the outskirts of and external to the CO disk. Based on their locations relative to the galaxy major/minor axes, as well as assumptions about the location of the RPS tail along the line-of-sight, we can infer what these residual velocities correspond to. These include:
    \begin{enumerate}
        \item Blueshifted gas on the leading side of the galaxy, which is evidence for strong \textit{ongoing} ram pressure in the region of the galaxy most exposed to the wind.
        \item Blueshifted clumps north of the inner tail components, which are likely regions being accelerated faster than the more bound and shielded inner tail.
        \item Two prominent tail features in the inner tail region, whose velocity residuals suggest that  features are stripped spiral arm components at different stages of inner-tail evolution.
        \item Redshifted gas around the western disk outskirts and the base of the RP tail. This might be CO emission from molecular gas clouds that have been stripped out of the disk, but only into the inner tail region and are thus now falling back into the disk.
    \end{enumerate}
    

    \item \textbf{Agreement with simulations:} The observations of NGC 4858 agree with predictions from simulations of ram pressure stripped galaxies experiencing highly-inclined ram pressure winds. When experiencing a highly edge-on wind, the evolution of gas density in the RPS tail should result in an asymmetric tail. Ultimately, inner-tail gas will coalesce on the side of the galaxy rotating into the RP wind where fallback is most likely to occur.
    
    This is precisely what is seen in the CO distribution of NGC 4858, with over 90\% of the CO gas emission beyond the gas disk truncation radius belonging to the Q4 quadrant. This quadrant is also the only region where significant fallback is detected, consistent with simulations such as \cite{zhu_when_2023}. 

    \item \textbf{Asymmetries in Disk Stellar Distributions:} We have found stellar complexes with head-tail morphologies in the stellar disk of NGC 4858, located only on the side of the galaxy rotating into the wind. These stellar complexes are associated with the CO emission in the disk. These stellar complexes are likely created from gas clouds which only form head-tail distributions in regions on the west side of the galaxy where  local ram pressure is stronger. These distributions are located near the edge of the stripped gas disk, so the local ram pressure would be approaching a maximum in this area.

    \item \textbf{Role of Ram Pressure Torques:} We emphasize the importance of torques due to ram pressure in highly inclined interactions. These torques help drive gas outwards on the side rotating with the wind, and help drive gas inwards on the side rotating against the wind. This contributes to the formation of asymmetric inner tails in ram pressure stripped galaxies, with excess dense gas on the side rotating against the wind. RP-induced torques driving gas towards the galaxy center may be important in explaining enhanced AGN activity in RPS galaxies \citep[e.g.][]{radovich_gasp_2019}.
\end{enumerate}

\newpage

\begin{acknowledgments}

We thank the anonymous referee for their comments and suggestions which helped improve the quality of the paper. We also thank Stephanie Tonnesen, Greg Bryan, Jingyao Zhu, and Cheng-Han Hsieh for insightful discussions.

This paper makes use of the following ALMA data: ADS/JAO.ALMA\#2021.1.01616.L. ALMA is a partnership of ESO (representing its member states), NSF (USA) and NINS (Japan), together with NRC (Canada), MOST and ASIAA (Taiwan), and KASI (Republic of Korea), in cooperation with the Republic of Chile. The Joint ALMA Observatory is operated by ESO, AUI/NRAO and NAOJ. The National Radio Astronomy Observatory is a facility of the National Science Foundation operated under cooperative agreement by Associated Universities, Inc.

This research is based in part on data collected at the Subaru Telescope, which is operated by the National Astronomical Observatory of Japan. We are honored and grateful for the opportunity of observing the Universe from Maunakea, which has the cultural, historical, and natural significance in Hawaii.

This paper also makes use of data based on observations made with the NASA/ESA Hubble Space Telescope, and obtained from the Hubble Legacy Archive, which is a collaboration between the Space Telescope Science Institute (STScI/NASA), the Space Telescope European Coordinating Facility (ST-ECF/ESA) and the Canadian Astronomy Data Centre (CADC/NRC/CSA).

HS acknowledges the support from the NRAO Student Observing Support (SOS) grant SOSPADA-031. 
PJ  acknowledges support from the project 25-19512L of the Czech Science Foundation and the institutional project RVO:67985815. PJ and BD further acknowledge support with observational ALMA data processing from the project LM2023059 of the Ministry of Education, Youth and Sports of the Czech Republic.
BD acknowledges funding from the HTM (grant TK202) and the EU Horizon Europe (EXCOSM, grant No. 101159513)."
MS acknowledges the support from the NSF grant 2407821. 
LC acknowledges support from the Australian Research Council’s Discovery Project funding scheme (DP210100337).
TS (DOI 10.54499/DL57/2016/CP1364/CT0009) is supported by national funds through Funda\c{c}\~{a}o para a Ci\^{e}ncia e a Tecnologia (FCT) and the Centro de Astrof\'isica da Universidade do Porto (CAUP).

\end{acknowledgments}

\newpage

\vspace{5mm}
\facilities{ALMA, Subaru(Suprime-Cam), HST(WFC)}


\software{CASA \citep{casa_team_casa_2022}, PHANGS-ALMA Pipeline \citep{leroy_phangsalma_2021}, Astropy, \textsc{Maskmoment},\citep{astropy_collaboration_astropy_2022}, Photutils \citep{larry_bradley_2024_12585239}, \Barolo, \citep{di_teodoro_3d_2015}}



\appendix

\section{Disk-Wind Angle Estimate}
\label{app:disk-wind-angle}

\begin{figure*}
    \centering

    \includegraphics[width=0.54\textwidth]{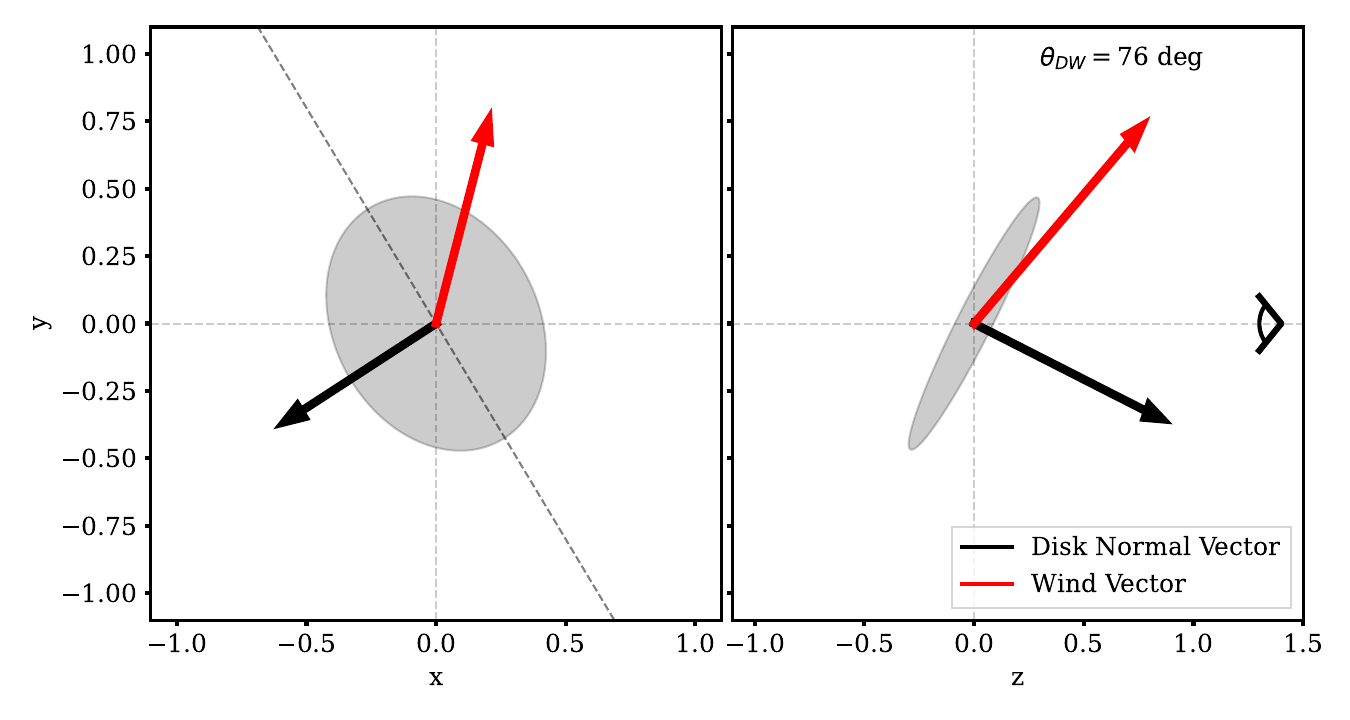}
    \includegraphics[width=0.45\textwidth]{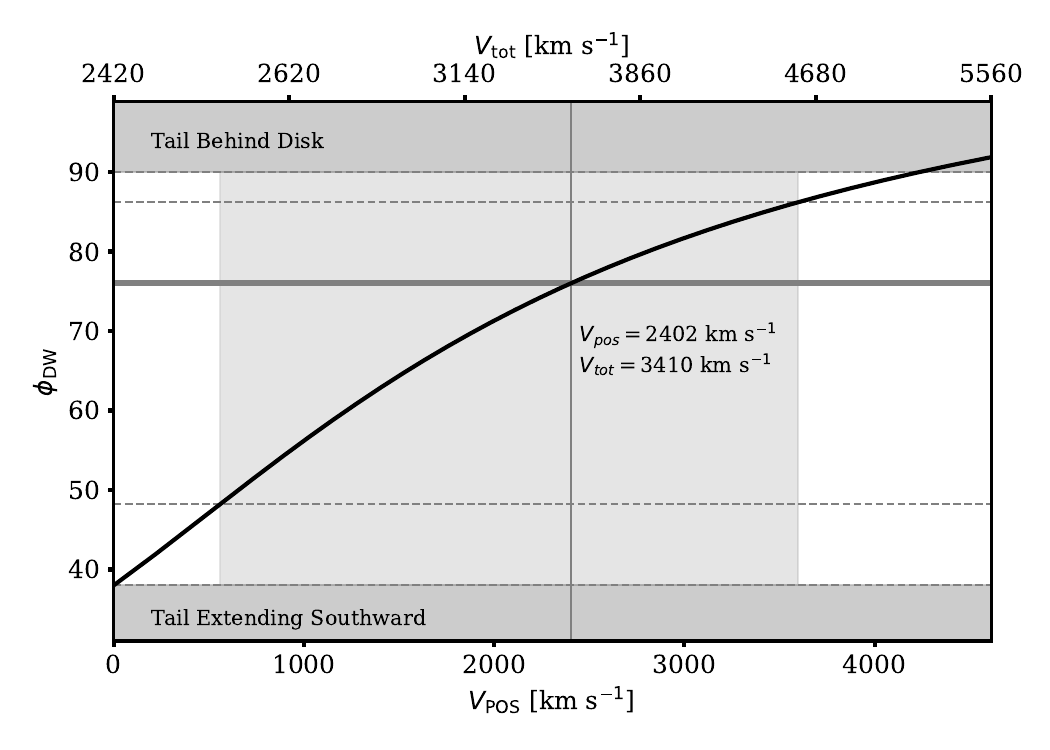}
    
    \caption{An overview of the NGC 4858 ram pressure interaction. \textbf{Left}: Projections of NGC 4858's stellar disk normal vector (black) and estimated RPS wind (red). Grey ellipses indicate the stellar disk position in each projection.  \textbf{Right}: Allowable disk-wind angles (and plane-of-sky velocities), given the observed tail geometry.}
    \label{fig:tail-geometry-details}
\end{figure*}

We can relate the physical disk-wind angle $\phi_\text{DW}$ to estimated projected angles ($\theta_{maj}$ and $\theta_{tail}$) and the estimated total galaxy velocity. The disk-wind angle is the angle between the wind vector and the normal vector to the disk plane, which is obtained by taking their dot product:

\begin{equation}
    \cos\left(\phi_{DW}\right) = \frac{\Vec{W}\cdot\Vec{n}}{|\Vec{W}||\Vec{n}|} 
\end{equation}

3D vectors for both the wind and the disk normal vector are required to fully describe the disk-wind angle. The normal vector for a disk inclined at an angle $i$ with a major axis position angle $\theta$ is given by

\begin{equation}
    \vec{n} = \begin{cases}
        \sin(i) \cos(\theta_\text{maj}) & \Hat{x} \\
        \sin(i) \sin(\theta_\text{maj}) & \Hat{y}\\
        \cos(i) & \Hat{z}
    \end{cases}
\end{equation}

The wind's 3D vector is related to the radial and plane-of-sky velocities. It is given by

\begin{equation}
    \vec{W} = \begin{cases}
        - V_\text{pos} \sin(\theta_\text{tail}) & \Hat{x}\\
        V_\text{pos} \cos(\theta_\text{tail}) & \Hat{y}\\
        V_\text{rad} & \Hat{z}
    \end{cases}
\end{equation}

where $\theta_\text{tail}$ is the wind angle in the plane-of-sky. By introducing $\theta_* = \theta_{\text{tail}} - \theta_\text{maj}$, we can define an expression for the disk-wind angle as

\begin{equation}
    \cos(\phi_\text{DW}) = \frac{V_\text{rad}}{V_\text{tot}}\cos(i) - \frac{V_\text{pos}}{V_\text{tot}}\sin(\theta_*)\sin(i) 
\end{equation}

where the total velocity is related to the radial and plane-of-sky velocities:
\begin{equation}
V_\text{tot} = \sqrt{V_{rad}^2 + V_{pos}^2}    
\end{equation}

Therefore, an estimate of the disk-wind angle can be made so long as  (1): the inclination and position angle of the stellar disk can be measured, (2): the near side of the disk can be determined, (3): the on-sky tail angle can be measured, and (4): a total orbital velocity can be estimated. Figure \ref{fig:tail-geometry-details} shows the results of this computation for NGC 4858.

\section{Features with High Velocity Residuals}
\label{app:high-velocity-residuals}

Under a cylindrical coordinate system, the possible velocity components are vertical velocities ($\Dot{z}$), radial velocities ($\Dot{r}$), and azimuthal velocities ($r\Dot{\phi}$). We will now describe when each of these velocities may correspond to observed velocity residuals in the galaxy. 

Vertical velocities have no azimuthal dependence, and can contribute anywhere along the plane of the disk. However, the vertical location of the gas is important in terms of whether blueshifted/redshifted motions correspond to motion vertically inward (vertical fallback) or outward (vertical stripping). For gas on the far side of the disk relative to the observer, blueshifted motions correspond to inward vertical motions, and vice versa for redshifted motion. The opposite is true for gas located on the near side of the galaxy relative to the observer. Since we conclude that the stripped galaxy tail extends towards the observer, most gas extending vertically above/below the gas disk should be nearer to the observer. Thus, any component of gas stripping in the vertical direction should correspond to blueshifted velocity residuals, and any gas falling back should create redshifted residuals.

Both radial motions and azimuthal motions have an azimuthal dependence, and their contribution to velocity residuals varies depending on their location in the disk plane relative to the galaxy major/minor axis. Along the galaxy major axis, azimuthal motions (gas travelling faster/slower than the rotational velocity predicted by the velocity model's rotation curve) can contribute to velocity residuals. Gas traveling faster than the expected circular velocity from the galaxy at that given radius will be blueshifted on the approaching side of the galaxy, and redshifted on the receding side of the galaxy. The opposite is true for redshifted gas.

Likewise, radial motions can contribute along the galaxy's minor axis. If a gas cloud along the minor axis is located on the near side of the galaxy relative to the observer, blueshifted motions will correspond to radial outward motion, and inward motion on the far side of the galaxy. The opposite is true for redshifted motions. 

Overall, velocity residuals will mostly be a combination of these possible sources. Along the major axis, it can be a combination of azimuthal/vertical velocities, and radial/vertical velocities along the minor axis. In an intermediate region, it can be a combination of all three.

\section{Effects of Changing the Rotation Curve on Velocity Residuals}
\label{app:changec-velresids}

\begin{figure*}
    \centering
    \includegraphics[width=0.9\textwidth]{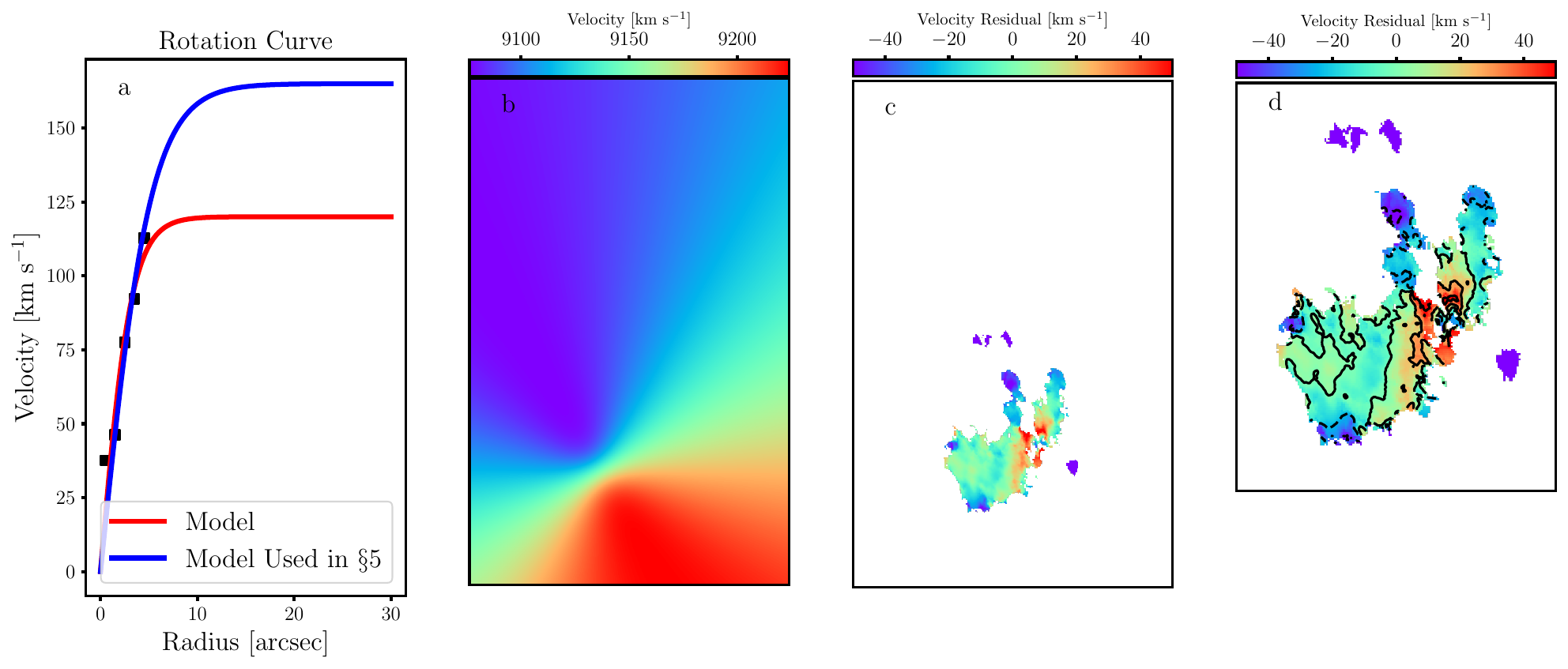}
    \caption{The effects of applying a rotation curve that immediately flattens beyond the CO truncation radius on the velocity residuals. a) A comparison of the two rotation curves. b) The derived velocity model. c) The velocity residuals of the galaxy. d) The velocity residuals, zoomed to show the main galaxy body, with overlaid isovelocity contours. }
    \label{app:velresid-differentcurve-1}
\end{figure*}

\begin{figure*}
    \centering
    \includegraphics[width=0.9\textwidth]{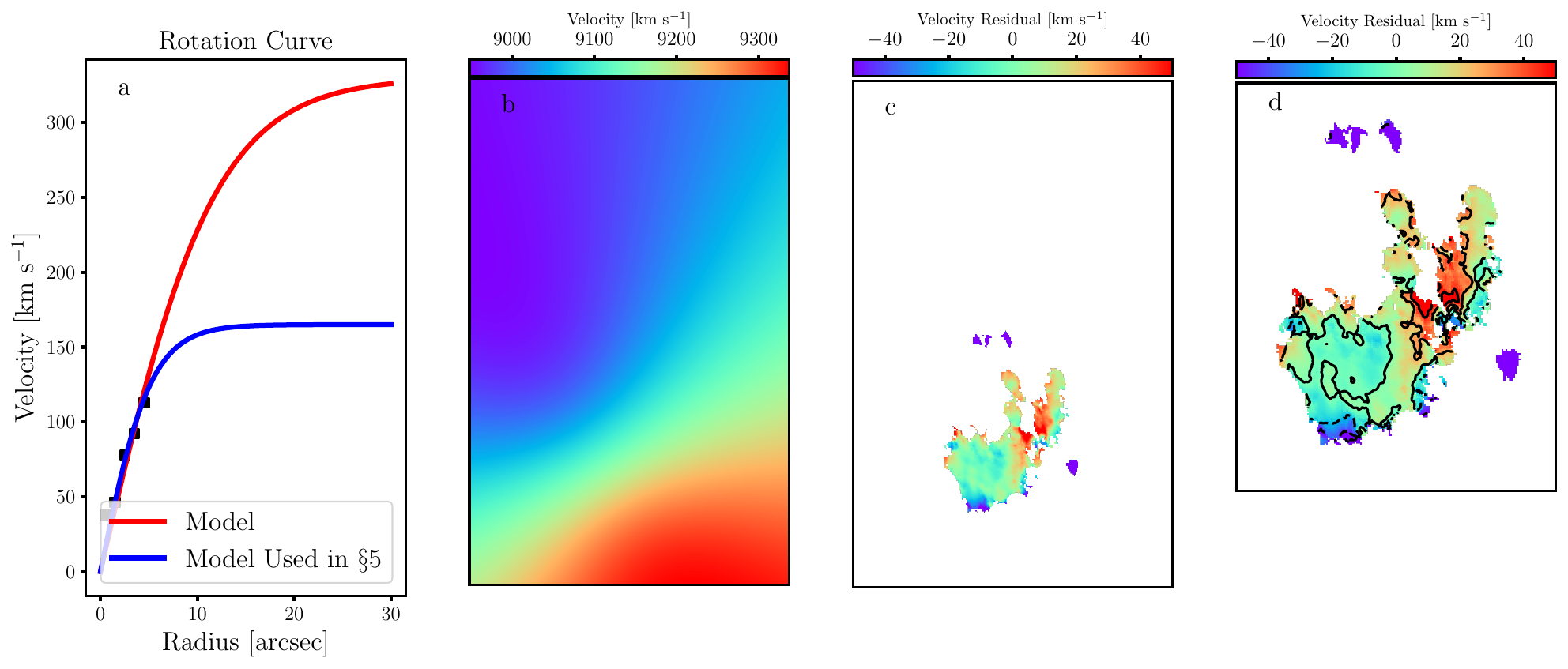}
    \caption{The same as Figure \ref{app:velresid-differentcurve-1}, but for a curve that has a $V_\text{max}$ of twice the value predicted from the Tully-Fisher relation.}
    \label{app:velresid-differentcurve-2}
\end{figure*}

In Figures \ref{app:velresid-differentcurve-1} and \ref{app:velresid-differentcurve-2} we show velocity residual maps after changing the input rotation curve. The two curves we test are a curve that immediately flattens after the CO truncation radius, and a curve that reaches twice the $V_\text{max}$ predicted from the Tully-Fisher relation. These Figures show that the redshifted region at the base of the two inner tail arms is robust to large changes in the rotation curve. Furthermore, a changing rotation curve cannot explain most blueshifted velocity residuals we believe are due to ram pressure.

\bibliography{references}{}

\begin{thebibliography}{}
\expandafter\ifx\csname natexlab\endcsname\relax\def\natexlab#1{#1}\fi
\providecommand{\url}[1]{\href{#1}{#1}}
\providecommand{\dodoi}[1]{doi:~\href{http://doi.org/#1}{\nolinkurl{#1}}}
\providecommand{\doeprint}[1]{\href{http://ascl.net/#1}{\nolinkurl{http://ascl.net/#1}}}
\providecommand{\doarXiv}[1]{\href{https://arxiv.org/abs/#1}{\nolinkurl{https://arxiv.org/abs/#1}}}

\bibitem[{Abadi {et~al.}(1999)Abadi, Moore, \& Bower}]{abadi_ram_1999}
Abadi, M.~G., Moore, B., \& Bower, R.~G. 1999, Monthly Notices of the Royal
  Astronomical Society, 308, 947, \dodoi{10.1046/j.1365-8711.1999.02715.x}

\bibitem[{Abramson \& Kenney(2014)}]{abramson_hubble_2014}
Abramson, A., \& Kenney, J. D.~P. 2014, The Astronomical Journal, 147, 63,
  \dodoi{10.1088/0004-6256/147/3/63}

\bibitem[{Akerman {et~al.}(2023)Akerman, Tonnesen, Poggianti, Smith, \&
  Marasco}]{akerman_how_2023}
Akerman, N., Tonnesen, S., Poggianti, B.~M., Smith, R., \& Marasco, A. 2023,
  The Astrophysical Journal, 948, 18, \dodoi{10.3847/1538-4357/acbf4d}

\bibitem[{Alam {et~al.}(2015)Alam, Albareti, Prieto, Anders, Anderson,
  Anderton, Andrews, Armengaud, Aubourg, Bailey, Basu, Bautista, Beaton, Beers,
  Bender, Berlind, Beutler, Bhardwaj, Bird, Bizyaev, Blake, Blanton, Blomqvist,
  Bochanski, Bolton, Bovy, Bradley, Brandt, Brauer, Brinkmann, Brown,
  Brownstein, Burden, Burtin, Busca, Cai, Capozzi, Rosell, Carr, Carrera,
  Chambers, Chaplin, Chen, Chiappini, Chojnowski, Chuang, Clerc, Comparat,
  Covey, Croft, Cuesta, Cunha, Costa, Rio, Davenport, Dawson, Lee, Delubac,
  Deshpande, Dhital, Dutra-Ferreira, Dwelly, Ealet, Ebelke, Edmondson,
  Eisenstein, Ellsworth, Elsworth, Epstein, Eracleous, Escoffier, Esposito,
  Evans, Fan, Fernández-Alvar, Feuillet, Ak, Finley, Finoguenov, Flaherty,
  Fleming, Font-Ribera, Foster, Frinchaboy, Galbraith-Frew, García,
  García-Hernández, Pérez, Gaulme, Ge, Génova-Santos, Georgakakis, Ghezzi,
  Gillespie, Girardi, Goddard, Gontcho, Hernández, Grebel, Green, Grieb,
  Grieves, Gunn, Guo, Harding, Hasselquist, Hawley, Hayden, Hearty, Hekker, Ho,
  Hogg, Holley-Bockelmann, Holtzman, Honscheid, Huber, Huehnerhoff, Ivans,
  Jiang, Johnson, Kinemuchi, Kirkby, Kitaura, Klaene, Knapp, Kneib, Koenig,
  Lam, Lan, Lang, Laurent, Goff, Leauthaud, Lee, Lee, Licquia, Liu, Long,
  López-Corredoira, Lorenzo-Oliveira, Lucatello, Lundgren, Lupton, III,
  Mahadevan, Maia, Majewski, Malanushenko, Malanushenko, Manchado, Manera, Mao,
  Maraston, Marchwinski, Margala, Martell, Martig, Masters, Mathur, McBride,
  McGehee, McGreer, McMahon, Ménard, Menzel, Merloni, Mészáros, Miller,
  Miralda-Escudé, Miyatake, Montero-Dorta, More, Morganson, Morice-Atkinson,
  Morrison, Mosser, Muna, Myers, Nandra, Newman, Neyrinck, Nguyen, Nichol,
  Nidever, Noterdaeme, Nuza, O’Connell, O’Connell, O’Connell, Ogando,
  Olmstead, Oravetz, Oravetz, Osumi, Owen, Padgett, Padmanabhan, Paegert,
  Palanque-Delabrouille, Pan, Parejko, Pâris, Park, Pattarakijwanich,
  Pellejero-Ibanez, Pepper, Percival, Pérez-Fournon, Pe´rez-Ra`fols,
  Petitjean, Pieri, Pinsonneault, Mello, Prada, Prakash, Price-Whelan,
  Protopapas, Raddick, Rahman, Reid, Rich, Rix, Robin, Rockosi, Rodrigues,
  Rodríguez-Torres, Roe, Ross, Ross, Rossi, Ruan, Rubiño-Martín, Rykoff,
  Salazar-Albornoz, Salvato, Samushia, Sánchez, Santiago, Sayres, Schiavon,
  Schlegel, Schmidt, Schneider, Schultheis, Schwope, Scóccola, Scott,
  Sellgren, Seo, Serenelli, Shane, Shen, Shetrone, Shu, Aguirre, Sivarani,
  Skrutskie, Slosar, Smith, Sobreira, Souto, Stassun, Steinmetz, Stello,
  Strauss, Streblyanska, Suzuki, Swanson, Tan, Tayar, Terrien, Thakar, Thomas,
  Thomas, Thompson, Tinker, Tojeiro, Troup, Vargas-Magaña, Vazquez, Verde,
  Viel, Vogt, Wake, Wang, Weaver, Weinberg, Weiner, White, Wilson, Wisniewski,
  Wood-Vasey, Ye`che, York, Zakamska, Zamora, Zasowski, Zehavi, Zhao, Zheng,
  Zhou~(周旭), Zhou~(周志民), Zou~(邹虎), \& Zhu}]{alam_eleventh_2015}
Alam, S., Albareti, F.~D., Prieto, C.~A., {et~al.} 2015, The Astrophysical
  Journal Supplement Series, 219, 12, \dodoi{10.1088/0067-0049/219/1/12}

\bibitem[{{Astropy Collaboration} {et~al.}(2022){Astropy Collaboration},
  Price-Whelan, Lim, Earl, Starkman, Bradley, Shupe, Patil, Corrales, Brasseur,
  Nöthe, Donath, Tollerud, Morris, Ginsburg, Vaher, Weaver, Tocknell,
  Jamieson, van Kerkwijk, Robitaille, Merry, Bachetti, Günther, Aldcroft,
  Alvarado-Montes, Archibald, Bódi, Bapat, Barentsen, Bazán, Biswas, Boquien,
  Burke, Cara, Cara, Conroy, Conseil, Craig, Cross, Cruz, D'Eugenio, Dencheva,
  Devillepoix, Dietrich, Eigenbrot, Erben, Ferreira, Foreman-Mackey, Fox,
  Freij, Garg, Geda, Glattly, Gondhalekar, Gordon, Grant, Greenfield, Groener,
  Guest, Gurovich, Handberg, Hart, Hatfield-Dodds, Homeier, Hosseinzadeh,
  Jenness, Jones, Joseph, Kalmbach, Karamehmetoglu, Kałuszyński, Kelley,
  Kern, Kerzendorf, Koch, Kulumani, Lee, Ly, Ma, MacBride, Maljaars, Muna,
  Murphy, Norman, O'Steen, Oman, Pacifici, Pascual, Pascual-Granado, Patil,
  Perren, Pickering, Rastogi, Roulston, Ryan, Rykoff, Sabater, Sakurikar,
  Salgado, Sanghi, Saunders, Savchenko, Schwardt, Seifert-Eckert, Shih, Jain,
  Shukla, Sick, Simpson, Singanamalla, Singer, Singhal, Sinha, Sipőcz,
  Spitler, Stansby, Streicher, Šumak, Swinbank, Taranu, Tewary, Tremblay,
  de~Val-Borro, Van~Kooten, Vasović, Verma, de~Miranda~Cardoso, Williams,
  Wilson, Winkel, Wood-Vasey, Xue, Yoachim, Zhang, Zonca, \& {Astropy Project
  Contributors}}]{astropy_collaboration_astropy_2022}
{Astropy Collaboration}, Price-Whelan, A.~M., Lim, P.~L., {et~al.} 2022, The
  Astrophysical Journal, 935, 167, \dodoi{10.3847/1538-4357/ac7c74}

\bibitem[{Bolatto {et~al.}(2013)Bolatto, Wolfire, \&
  Leroy}]{bolatto_co--h2_2013}
Bolatto, A.~D., Wolfire, M., \& Leroy, A.~K. 2013, Annual Review of Astronomy
  and Astrophysics, 51, 207, \dodoi{10.1146/annurev-astro-082812-140944}

\bibitem[{Boselli {et~al.}(2022)Boselli, Fossati, \& Sun}]{boselli_ram_2022}
Boselli, A., Fossati, M., \& Sun, M. 2022, The Astronomy and Astrophysics
  Review, 30, 3, \dodoi{10.1007/s00159-022-00140-3}

\bibitem[{Boselli \& Gavazzi(2006)}]{boselli_environmental_2006}
Boselli, A., \& Gavazzi, G. 2006, Publications of the Astronomical Society of
  the Pacific, 118, 517, \dodoi{10.1086/500691}

\bibitem[{Boselli {et~al.}(2016)Boselli, Cuillandre, Fossati, Boissier, Bomans,
  Consolandi, Anselmi, Cortese, Côté, Durrell, Ferrarese, Fumagalli, Gavazzi,
  Gwyn, Hensler, Sun, \& Toloba}]{boselli_spectacular_2016}
Boselli, A., Cuillandre, J.~C., Fossati, M., {et~al.} 2016, Astronomy and
  Astrophysics, 587, A68, \dodoi{10.1051/0004-6361/201527795}

\bibitem[{Boselli {et~al.}(2021)Boselli, Lupi, Epinat, Amram, Fossati,
  Anderson, Boissier, Boquien, Consolandi, Côté, Cuillandre, Ferrarese,
  Galbany, Gavazzi, Gómez-López, Gwyn, Hensler, Hutchings, Kuncarayakti,
  Longobardi, Peng, Plana, Postma, Roediger, Roehlly, Schimd, Trinchieri, \&
  Vollmer}]{boselli_virgo_2021}
Boselli, A., Lupi, A., Epinat, B., {et~al.} 2021, Astronomy \& Astrophysics,
  646, A139, \dodoi{10.1051/0004-6361/202039046}

\bibitem[{Bradley {et~al.}(2024)Bradley, Sipőcz, Robitaille, Tollerud,
  Vinícius, Deil, Barbary, Wilson, Busko, Donath, Günther, Cara, Lim,
  Meßlinger, Burnett, Conseil, Droettboom, Bostroem, Bray, Bratholm, Jamieson,
  Ginsburg, Barentsen, Craig, Pascual, Rathi, Perrin, Morris, \&
  Perren}]{larry_bradley_2024_12585239}
Bradley, L., Sipőcz, B., Robitaille, T., {et~al.} 2024, astropy/photutils:
  1.13.0,  Zenodo, \dodoi{10.5281/zenodo.12585239}

\bibitem[{Brown {et~al.}(2023)Brown, Roberts, Thorp, Ellison, Zabel, Wilson,
  Bahé, Bisaria, Bolatto, Boselli, Chung, Cortese, Catinella, Davis,
  Jiménez-Donaire, Lagos, Lee, Parker, Smith, Spekkens, Stevens, Villanueva,
  \& Watts}]{brown_vertico_2023}
Brown, T., Roberts, I.~D., Thorp, M., {et~al.} 2023, The Astrophysical Journal,
  956, 37, \dodoi{10.3847/1538-4357/acf195}

\bibitem[{Cayatte {et~al.}(1990)Cayatte, van Gorkom, Balkowski, \&
  Kotanyi}]{cayatte_vla_1990}
Cayatte, V., van Gorkom, J.~H., Balkowski, C., \& Kotanyi, C. 1990, The
  Astronomical Journal, 100, 604, \dodoi{10.1086/115545}

\bibitem[{Chen {et~al.}(2020)Chen, Sun, Yagi, Bravo-Alfaro, Brinks, Kenney,
  Combes, Sivanandam, Jachym, Fossati, Gavazzi, Boselli, Nulsen, Sarazin, Ge,
  Yoshida, \& Roediger}]{chen_ram_2020}
Chen, H., Sun, M., Yagi, M., {et~al.} 2020, Monthly Notices of the Royal
  Astronomical Society, 496, 4654, \dodoi{10.1093/mnras/staa1868}

\bibitem[{Chung {et~al.}(2009)Chung, van Gorkom, Kenney, Crowl, \&
  Vollmer}]{chung_vla_2009}
Chung, A., van Gorkom, J.~H., Kenney, J. D.~P., Crowl, H., \& Vollmer, B. 2009,
  The Astronomical Journal, 138, 1741, \dodoi{10.1088/0004-6256/138/6/1741}

\bibitem[{Cortese {et~al.}(2021)Cortese, Catinella, \&
  Smith}]{cortese_dawes_2021}
Cortese, L., Catinella, B., \& Smith, R. 2021, Publications of the Astronomical
  Society of Australia, 38, e035, \dodoi{10.1017/pasa.2021.18}

\bibitem[{Cramer {et~al.}(2020)Cramer, Kenney, Cortes, Cortes, Vlahakis,
  Jáchym, Pompei, \& Rubio}]{cramer_alma_2020}
Cramer, W.~J., Kenney, J. D.~P., Cortes, J.~R., {et~al.} 2020, The
  Astrophysical Journal, 901, 95, \dodoi{10.3847/1538-4357/abaf54}

\bibitem[{Cramer {et~al.}(2019)Cramer, Kenney, Sun, Crowl, Yagi, Jáchym,
  Roediger, \& Waldron}]{cramer_spectacular_2019}
Cramer, W.~J., Kenney, J. D.~P., Sun, M., {et~al.} 2019, The Astrophysical
  Journal, 870, 63, \dodoi{10.3847/1538-4357/aaefff}

\bibitem[{Cramer {et~al.}(2021)Cramer, Kenney, Tonnesen, Smith, Wong, Jáchym,
  Cortés, Cortés, \& Wu}]{cramer_molecular_2021}
Cramer, W.~J., Kenney, J. D.~P., Tonnesen, S., {et~al.} 2021, The Astrophysical
  Journal, 921, 22, \dodoi{10.3847/1538-4357/ac1793}

\bibitem[{Di~Teodoro \& Fraternali(2015)}]{di_teodoro_3d_2015}
Di~Teodoro, E.~M., \& Fraternali, F. 2015, Monthly Notices of the Royal
  Astronomical Society, 451, 3021, \dodoi{10.1093/mnras/stv1213}

\bibitem[{Ebeling {et~al.}(2014)Ebeling, Stephenson, \&
  Edge}]{ebeling_jellyfish_2014}
Ebeling, H., Stephenson, L.~N., \& Edge, A.~C. 2014, The Astrophysical Journal,
  781, L40, \dodoi{10.1088/2041-8205/781/2/L40}

\bibitem[{Fossati {et~al.}(2012)Fossati, Gavazzi, Boselli, \&
  Fumagalli}]{fossati_65_2012}
Fossati, M., Gavazzi, G., Boselli, A., \& Fumagalli, M. 2012, Astronomy \&
  Astrophysics, 544, A128, \dodoi{10.1051/0004-6361/201219933}

\bibitem[{Gallego {et~al.}(1996)Gallego, Zamorano, Rego, Alonso, \&
  Vitores}]{gallego_observations_1996}
Gallego, J., Zamorano, J., Rego, M., Alonso, O., \& Vitores, A.~G. 1996,
  Astronomy and Astrophysics Supplement Series, 120, 323,
  \dodoi{10.1051/aas:1996297}

\bibitem[{Gavazzi {et~al.}(2001)Gavazzi, Boselli, Mayer, Iglesias-Paramo,
  Vílchez, \& Carrasco}]{gavazzi_75_2001}
Gavazzi, G., Boselli, A., Mayer, L., {et~al.} 2001, The Astrophysical Journal,
  563, L23, \dodoi{10.1086/338389}

\bibitem[{Gavazzi {et~al.}(2018)Gavazzi, Consolandi, Gutierrez, Boselli, \&
  Yoshida}]{gavazzi_ubiquitous_2018}
Gavazzi, G., Consolandi, G., Gutierrez, M.~L., Boselli, A., \& Yoshida, M.
  2018, Astronomy and Astrophysics, 618, A130,
  \dodoi{10.1051/0004-6361/201833427}

\bibitem[{Geha {et~al.}(2012)Geha, Blanton, Yan, \& Tinker}]{geha_slar_2012}
Geha, M., Blanton, M.~R., Yan, R., \& Tinker, J.~L. 2012, The Astrophysical
  Journal, 757, 85, \dodoi{10.1088/0004-637X/757/1/85}

\bibitem[{George {et~al.}(2023)George, Poggianti, Tomičić, Postma, Côté,
  Fritz, Ghosh, Gullieuszik, Hutchings, Moretti, Omizzolo, Radovich, Sreekumar,
  Subramaniam, Tandon, \& Vulcani}]{george_ultraviolet_2023}
George, K., Poggianti, B.~M., Tomičić, N., {et~al.} 2023, Monthly Notices of
  the Royal Astronomical Society, 519, 2426, \dodoi{10.1093/mnras/stac3593}

\bibitem[{Giovanelli \& Haynes(1985)}]{giovanelli_gas_1985}
Giovanelli, R., \& Haynes, M.~P. 1985, The Astrophysical Journal, 292, 404,
  \dodoi{10.1086/163170}

\bibitem[{Gunn \& Gott(1972)}]{gunn_infall_1972}
Gunn, J.~E., \& Gott, III, J.~R. 1972, The Astrophysical Journal, 176, 1,
  \dodoi{10.1086/151605}

\bibitem[{Hinshaw {et~al.}(2013)Hinshaw, Larson, Komatsu, Spergel, Bennett,
  Dunkley, Nolta, Halpern, Hill, Odegard, Page, Smith, Weiland, Gold, Jarosik,
  Kogut, Limon, Meyer, Tucker, Wollack, \& Wright}]{hinshaw_nine-year_2013}
Hinshaw, G., Larson, D., Komatsu, E., {et~al.} 2013, The Astrophysical Journal
  Supplement Series, 208, 19, \dodoi{10.1088/0067-0049/208/2/19}

\bibitem[{Iye {et~al.}(2019)Iye, Tadaki, \& Fukumoto}]{iye_spin_2019}
Iye, M., Tadaki, K.-i., \& Fukumoto, H. 2019, The Astrophysical Journal, 886,
  133, \dodoi{10.3847/1538-4357/ab4a18}

\bibitem[{Jedrzejewski(1987)}]{jedrzejewski_ccd_1987}
Jedrzejewski, R.~I. 1987, Monthly Notices of the Royal Astronomical Society,
  226, 747, \dodoi{10.1093/mnras/226.4.747}

\bibitem[{Jian {et~al.}(2023)Jian, Lin, Hsieh, Umetsu, Lopez-Coba, Oguri,
  Bottrell, Toba, Koyama, Chang, Kodama, Komiyama, More, Lin, Nishizawa, \&
  Tanaka}]{jian_radial_2023}
Jian, H.-Y., Lin, L., Hsieh, B.-C., {et~al.} 2023, The Astrophysical Journal,
  957, 85, \dodoi{10.3847/1538-4357/acfc22}

\bibitem[{Jáchym {et~al.}(2009)Jáchym, Köppen, Palouš, \&
  Combes}]{jachym_ram_2009}
Jáchym, P., Köppen, J., Palouš, J., \& Combes, F. 2009, Astronomy and
  Astrophysics, 500, 693, \dodoi{10.1051/0004-6361/200811469}

\bibitem[{Jáchym {et~al.}(2019)Jáchym, Kenney, Sun, Combes, Cortese, Scott,
  Sivanandam, Brinks, Roediger, Palouš, \& Fumagalli}]{jachym_alma_2019}
Jáchym, P., Kenney, J. D.~P., Sun, M., {et~al.} 2019, The Astrophysical
  Journal, 883, 145, \dodoi{10.3847/1538-4357/ab3e6c}

\bibitem[{Kenney {et~al.}(2015)Kenney, Abramson, \&
  Bravo-Alfaro}]{kenney_hst_2015}
Kenney, J. D.~P., Abramson, A., \& Bravo-Alfaro, H. 2015, The Astronomical
  Journal, 150, 59, \dodoi{10.1088/0004-6256/150/2/59}

\bibitem[{Kenney \& Koopmann(1999)}]{kenney_ongoing_1999}
Kenney, J. D.~P., \& Koopmann, R.~A. 1999, The Astronomical Journal, 117, 181,
  \dodoi{10.1086/300683}

\bibitem[{Kenney {et~al.}(1992)Kenney, Wilson, Scoville, Devereux, \&
  Young}]{kenney_twin_1992}
Kenney, J. D.~P., Wilson, C.~D., Scoville, N.~Z., Devereux, N.~A., \& Young,
  J.~S. 1992, The Astrophysical Journal, 395, L79, \dodoi{10.1086/186492}

\bibitem[{Koopmann \& Kenney(2004)}]{koopmann_h_2004}
Koopmann, R.~A., \& Kenney, J. D.~P. 2004, The Astrophysical Journal, 613, 866,
  \dodoi{10.1086/423191}

\bibitem[{Köppen {et~al.}(2018)Köppen, Jáchym, Taylor, \&
  Palouš}]{koppen_ram_2018}
Köppen, J., Jáchym, P., Taylor, R., \& Palouš, J. 2018, Monthly Notices of
  the Royal Astronomical Society, 479, 4367, \dodoi{10.1093/mnras/sty1610}

\bibitem[{Lee {et~al.}(2020)Lee, Kimm, Katz, Rosdahl, Devriendt, \&
  Slyz}]{lee_dual_2020}
Lee, J., Kimm, T., Katz, H., {et~al.} 2020, The Astrophysical Journal, 905, 31,
  \dodoi{10.3847/1538-4357/abc3b8}

\bibitem[{Leroy {et~al.}(2009)Leroy, Walter, Bigiel, Usero, Weiss, Brinks,
  de~Blok, Kennicutt, Schuster, Kramer, Wiesemeyer, \&
  Roussel}]{leroy_heracles_2009}
Leroy, A.~K., Walter, F., Bigiel, F., {et~al.} 2009, The Astronomical Journal,
  137, 4670, \dodoi{10.1088/0004-6256/137/6/4670}

\bibitem[{Leroy {et~al.}(2021)Leroy, Hughes, Liu, Pety, Rosolowsky, Saito,
  Schinnerer, Schruba, Usero, Faesi, Herrera, Chevance, Hygate, Kepley, Koch,
  Querejeta, Sliwa, Will, Wilson, Anand, Barnes, Belfiore, Bešlić, Bigiel,
  Blanc, Bolatto, Boquien, Cao, Chandar, Chastenet, Chiang, Congiu, Dale,
  Deger, Brok, Eibensteiner, Emsellem, García-Rodríguez, Glover, Grasha,
  Groves, Henshaw, Donaire, Kim, Klessen, Kreckel, Kruijssen, Larson, Lee,
  Mayker, McElroy, Meidt, Mok, Pan, Puschnig, Razza, Sánchez-Bl’azquez,
  Sandstrom, Santoro, Sardone, Scheuermann, Sun, Thilker, Turner, Ubeda, Utomo,
  Watkins, \& Williams}]{leroy_phangsalma_2021}
Leroy, A.~K., Hughes, A., Liu, D., {et~al.} 2021, The Astrophysical Journal
  Supplement Series, 255, 19, \dodoi{10.3847/1538-4365/abec80}

\bibitem[{Lokas \& Mamon(2003)}]{lokas_dark_2003}
Lokas, E.~L., \& Mamon, G.~A. 2003, Monthly Notices of the Royal Astronomical
  Society, 343, 401, \dodoi{10.1046/j.1365-8711.2003.06684.x}

\bibitem[{McPartland {et~al.}(2016)McPartland, Ebeling, Roediger, \&
  Blumenthal}]{mcpartland_jellyfish_2016}
McPartland, C., Ebeling, H., Roediger, E., \& Blumenthal, K. 2016, Monthly
  Notices of the Royal Astronomical Society, 455, 2994,
  \dodoi{10.1093/mnras/stv2508}

\bibitem[{Miyazaki {et~al.}(2002)Miyazaki, Komiyama, Sekiguchi, Okamura, Doi,
  Furusawa, Hamabe, Imi, Kimura, Nakata, Okada, Ouchi, Shimasaku, Yagi, \&
  Yasuda}]{miyazaki_subaru_2002}
Miyazaki, S., Komiyama, Y., Sekiguchi, M., {et~al.} 2002, Publications of the
  Astronomical Society of Japan, 54, 833, \dodoi{10.1093/pasj/54.6.833}

\bibitem[{Molnár {et~al.}(2022)Molnár, Serra, van~der Hulst, Jarrett,
  Boselli, Cortese, Healy, de~Blok, Cappellari, Hess, Józsa, McDermid,
  Oosterloo, \& Verheijen}]{molnar_westerbork_2022}
Molnár, D.~C., Serra, P., van~der Hulst, T., {et~al.} 2022, Astronomy and
  Astrophysics, 659, A94, \dodoi{10.1051/0004-6361/202142614}

\bibitem[{Moretti {et~al.}(2020)Moretti, Paladino, Poggianti, Serra, Ramatsoku,
  Franchetto, Deb, Gullieuszik, Tomičić, Mingozzi, Vulcani, Radovich,
  Bettoni, \& Fritz}]{moretti_high_2020}
Moretti, A., Paladino, R., Poggianti, B.~M., {et~al.} 2020, The Astrophysical
  Journal, 897, L30, \dodoi{10.3847/2041-8213/ab9f3b}

\bibitem[{Parkash {et~al.}(2018)Parkash, Brown, Jarrett, \&
  Bonne}]{parkash_relationships_2018}
Parkash, V., Brown, M. J.~I., Jarrett, T.~H., \& Bonne, N.~J. 2018, The
  Astrophysical Journal, 864, 40, \dodoi{10.3847/1538-4357/aad3b9}

\bibitem[{Poggianti {et~al.}(2017)Poggianti, Moretti, Gullieuszik, Fritz,
  Jaffé, Bettoni, Fasano, Bellhouse, Hau, Vulcani, Biviano, Omizzolo,
  Paccagnella, D'Onofrio, Cava, Sheen, Couch, \& Owers}]{poggianti_gasp_2017}
Poggianti, B.~M., Moretti, A., Gullieuszik, M., {et~al.} 2017, The
  Astrophysical Journal, 844, 48, \dodoi{10.3847/1538-4357/aa78ed}

\bibitem[{Poggianti {et~al.}(2019)Poggianti, Gullieuszik, Tonnesen, Moretti,
  Vulcani, Radovich, Jaffé, Fritz, Bettoni, Franchetto, Fasano, Bellhouse, \&
  Omizzolo}]{poggianti_gasp_2019-1}
Poggianti, B.~M., Gullieuszik, M., Tonnesen, S., {et~al.} 2019, Monthly Notices
  of the Royal Astronomical Society, 482, 4466, \dodoi{10.1093/mnras/sty2999}

\bibitem[{Popping {et~al.}(2014)Popping, Somerville, \&
  Trager}]{popping_evolution_2014}
Popping, G., Somerville, R.~S., \& Trager, S.~C. 2014, Monthly Notices of the
  Royal Astronomical Society, 442, 2398, \dodoi{10.1093/mnras/stu991}

\bibitem[{Price-Whelan {et~al.}(2024)Price-Whelan, Wagg, Sipőcz, Lilleengen,
  Starkman, NGC, Lenz, Greco, Hart, Robert, Foreman-Mackey, HNLala, Lim, Oh,
  Koposov, \& Li}]{price-whelan_adrngala_2024}
Price-Whelan, A., Wagg, T., Sipőcz, B., {et~al.} 2024, adrn/gala: v1.8.1,
  Zenodo, \dodoi{10.5281/zenodo.10449846}

\bibitem[{Radovich {et~al.}(2019)Radovich, Poggianti, Jaffé, Moretti, Bettoni,
  Gullieuszik, Vulcani, \& Fritz}]{radovich_gasp_2019}
Radovich, M., Poggianti, B., Jaffé, Y.~L., {et~al.} 2019, Monthly Notices of
  the Royal Astronomical Society, 486, 486, \dodoi{10.1093/mnras/stz809}

\bibitem[{Roberts {et~al.}(2021)Roberts, Weeren, McGee, Botteon, Ignesti, \&
  Rottgering}]{roberts_lotss_2021}
Roberts, I.~D., Weeren, R. J.~v., McGee, S.~L., {et~al.} 2021, Astronomy \&
  Astrophysics, 652, A153, \dodoi{10.1051/0004-6361/202141118}

\bibitem[{Roberts {et~al.}(2024)Roberts, van Weeren, Lal, Sun, Chen, Ignesti,
  Brüggen, Lyskova, Venturi, \& Yagi}]{roberts_radio-continuum_2024}
Roberts, I.~D., van Weeren, R.~J., Lal, D.~V., {et~al.} 2024, Astronomy and
  Astrophysics, 683, A11, \dodoi{10.1051/0004-6361/202347977}

\bibitem[{Roediger {et~al.}(2014)Roediger, Bruggen, Owers, Ebeling, \&
  Sun}]{roediger_star_2014}
Roediger, E., Bruggen, M., Owers, M.~S., Ebeling, H., \& Sun, M. 2014, Monthly
  Notices of the Royal Astronomical Society, 443, L114,
  \dodoi{10.1093/mnrasl/slu087}

\bibitem[{Roediger \& Brüggen(2006)}]{roediger_ram_2006}
Roediger, E., \& Brüggen, M. 2006, Monthly Notices of the Royal Astronomical
  Society, 369, 567, \dodoi{10.1111/j.1365-2966.2006.10335.x}

\bibitem[{Samuel {et~al.}(2023)Samuel, Pardasani, Wetzel, Santistevan,
  Boylan-Kolchin, Moreno, \& Faucher-Giguère}]{samuel_jolt_2023}
Samuel, J., Pardasani, B., Wetzel, A., {et~al.} 2023, Monthly Notices of the
  Royal Astronomical Society, 525, 3849, \dodoi{10.1093/mnras/stad2576}

\bibitem[{Sellwood \& Wilkinson(1993)}]{sellwood_dynamics_1993}
Sellwood, J.~A., \& Wilkinson, A. 1993, Reports on Progress in Physics, 56,
  173, \dodoi{10.1088/0034-4885/56/2/001}

\bibitem[{Shimwell {et~al.}(2017)Shimwell, Röttgering, Best, Williams,
  Dijkema, Gasperin, Hardcastle, Heald, Hoang, Horneffer, Intema, Mahony,
  Mandal, Mechev, Morabito, Oonk, Rafferty, Retana-Montenegro, Sabater, Tasse,
  Weeren, Brüggen, Brunetti, Chyży, Conway, Haverkorn, Jackson, Jarvis,
  McKean, Miley, Morganti, White, Wise, Bemmel, Beck, Brienza, Bonafede,
  Rivera, Cassano, Clarke, Cseh, Deller, Drabent, Driel, Engels, Falcke,
  Ferrari, Fröhlich, Garrett, Harwood, Heesen, Hoeft, Horellou, Israel,
  Kapińska, Kunert-Bajraszewska, McKay, Mohan, Orrú, Pizzo, Prandoni,
  Schwarz, Shulevski, Sipior, Smith, Sridhar, Steinmetz, Stroe, Varenius, Werf,
  Zensus, \& Zwart}]{shimwell_lofar_2017}
Shimwell, T.~W., Röttgering, H. J.~A., Best, P.~N., {et~al.} 2017, Astronomy
  \& Astrophysics, 598, A104, \dodoi{10.1051/0004-6361/201629313}

\bibitem[{Smith {et~al.}(2010)Smith, Lucey, Hammer, Hornschemeier, Carter,
  Hudson, Marzke, Mouhcine, Eftekharzadeh, James, Khosroshahi, Kourkchi, \&
  Karick}]{smith_ultraviolet_2010}
Smith, R.~J., Lucey, J.~R., Hammer, D., {et~al.} 2010, Monthly Notices of the
  Royal Astronomical Society, 408, 1417,
  \dodoi{10.1111/j.1365-2966.2010.17253.x}

\bibitem[{Sohn {et~al.}(2017)Sohn, Geller, Zahid, Fabricant, Diaferio, \&
  Rines}]{sohn_velocity_2017}
Sohn, J., Geller, M.~J., Zahid, H.~J., {et~al.} 2017, The Astrophysical Journal
  Supplement Series, 229, 20, \dodoi{10.3847/1538-4365/aa653e}

\bibitem[{Solanes {et~al.}(2001)Solanes, Manrique, García-Gómez,
  González-Casado, Giovanelli, \& Haynes}]{solanes_h_2001}
Solanes, J.~M., Manrique, A., García-Gómez, C., {et~al.} 2001, The
  Astrophysical Journal, 548, 97, \dodoi{10.1086/318672}

\bibitem[{Solomon {et~al.}(1997)Solomon, Downes, Radford, \&
  Barrett}]{solomon_molecular_1997}
Solomon, P.~M., Downes, D., Radford, S. J.~E., \& Barrett, J.~W. 1997, The
  Astrophysical Journal, 478, 144, \dodoi{10.1086/303765}

\bibitem[{Sparre {et~al.}(2023)Sparre, Pfrommer, \&
  Puchwein}]{sparre_magnetised_2023}
Sparre, M., Pfrommer, C., \& Puchwein, E. 2023, Monthly Notices of the Royal
  Astronomical Society, 527, 5829, \dodoi{10.1093/mnras/stad3607}

\bibitem[{Sun {et~al.}(2010)Sun, Donahue, Roediger, Nulsen, Voit, Sarazin,
  Forman, \& Jones}]{sun_spectacular_2010}
Sun, M., Donahue, M., Roediger, E., {et~al.} 2010, The Astrophysical Journal,
  708, 946, \dodoi{10.1088/0004-637X/708/2/946}

\bibitem[{Sun {et~al.}(2021)Sun, Ge, Luo, Yagi, Jáchym, Boselli, Fossati,
  Nulsen, Yoshida, \& Gavazzi}]{sun_universal_2021}
Sun, M., Ge, C., Luo, R., {et~al.} 2021, Nature Astronomy, 6, 270,
  \dodoi{10.1038/s41550-021-01516-8}

\bibitem[{Taylor {et~al.}(2020)Taylor, Köppen, Jáchym, Minchin, Palouš, \&
  Wünsch}]{taylor_faint_2020}
Taylor, R., Köppen, J., Jáchym, P., {et~al.} 2020, The Astronomical Journal,
  159, 218, \dodoi{10.3847/1538-3881/ab6988}

\bibitem[{Team {et~al.}(2022)Team, Bean, Bhatnagar, Castro, Meyer, Emonts,
  Garcia, Garwood, Golap, Villalba, Harris, Hayashi, Hoskins, Hsieh,
  Jagannathan, Kawasaki, Keimpema, Kettenis, Lopez, Marvil, Masters, McNichols,
  Mehringer, Miel, Moellenbrock, Montesino, Nakazato, Ott, Petry, Pokorny,
  Raba, Rau, Schiebel, Schweighart, Sekhar, Shimada, Small, Steeb, Sugimoto,
  Suoranta, Tsutsumi, Bemmel, Verkouter, Wells, Xiong, Szomoru, Griffith,
  Glendenning, \& Kern}]{casa_team_casa_2022}
Team, C., Bean, B., Bhatnagar, S., {et~al.} 2022, Publications of the
  Astronomical Society of the Pacific, 134, 114501,
  \dodoi{10.1088/1538-3873/ac9642}

\bibitem[{Tonnesen \& Bryan(2009)}]{tonnesen_gas_2009}
Tonnesen, S., \& Bryan, G.~L. 2009, The Astrophysical Journal, 694, 789,
  \dodoi{10.1088/0004-637X/694/2/789}

\bibitem[{Tonnesen \& Bryan(2012)}]{tonnesen_star_2012}
---. 2012, Monthly Notices of the Royal Astronomical Society, 422, 1609,
  \dodoi{10.1111/j.1365-2966.2012.20737.x}

\bibitem[{Tonnesen \& Bryan(2021)}]{tonnesen_its_2021}
---. 2021, The Astrophysical Journal, 911, 68, \dodoi{10.3847/1538-4357/abe7e2}

\bibitem[{Tonnesen {et~al.}(2007)Tonnesen, Bryan, \& van
  Gorkom}]{tonnesen_environmentally-driven_2007}
Tonnesen, S., Bryan, G.~L., \& van Gorkom, J.~H. 2007, Environmentally-{Driven}
  {Evolution} of {Simulated} {Cluster} {Galaxies}, \dodoi{10.1086/523034}

\bibitem[{Troncoso-Iribarren {et~al.}(2020)Troncoso-Iribarren, Padilla,
  Santander, Lagos, García-Lambas, Rodríguez, \&
  Contreras}]{troncoso-iribarren_better_2020}
Troncoso-Iribarren, P., Padilla, N., Santander, C., {et~al.} 2020, Monthly
  Notices of the Royal Astronomical Society, 497, 4145,
  \dodoi{10.1093/mnras/staa274}

\bibitem[{Tully \& Fisher(1977)}]{tully_new_1977}
Tully, R.~B., \& Fisher, J.~R. 1977, Astronomy and Astrophysics, 54, 661.
\newblock \url{https://ui.adsabs.harvard.edu/abs/1977A&A....54..661T}

\bibitem[{van~der Kruit \& Allen(1978)}]{van_der_kruit_kinematics_1978}
van~der Kruit, P.~C., \& Allen, R.~J. 1978, Annual Review of Astronomy and
  Astrophysics, 16, 103, \dodoi{10.1146/annurev.aa.16.090178.000535}

\bibitem[{Vollmer {et~al.}(2001)Vollmer, Cayatte, Balkowski, \&
  Duschl}]{vollmer_ram_2001}
Vollmer, B., Cayatte, V., Balkowski, C., \& Duschl, W.~J. 2001, The
  Astrophysical Journal, 561, 708, \dodoi{10.1086/323368}

\bibitem[{Vulcani {et~al.}(2024)Vulcani, Moretti, Poggianti, Radovich, Werle,
  Gullieuszik, Fritz, Bacchini, \& Richard}]{vulcani_evidence_2024}
Vulcani, B., Moretti, A., Poggianti, B.~M., {et~al.} 2024, Astronomy and
  Astrophysics, 682, A117, \dodoi{10.1051/0004-6361/202348135}

\bibitem[{Wetzel {et~al.}(2012)Wetzel, Tinker, \& Conroy}]{wetzel_galaxy_2012}
Wetzel, A.~R., Tinker, J.~L., \& Conroy, C. 2012, Monthly Notices of the Royal
  Astronomical Society, 424, 232, \dodoi{10.1111/j.1365-2966.2012.21188.x}

\bibitem[{Yagi {et~al.}(2017)Yagi, Yoshida, Gavazzi, Komiyama, Kashikawa, \&
  Okamura}]{yagi_extended_2017}
Yagi, M., Yoshida, M., Gavazzi, G., {et~al.} 2017, The Astrophysical Journal,
  839, 65, \dodoi{10.3847/1538-4357/aa68e3}

\bibitem[{Yagi {et~al.}(2010)Yagi, Yoshida, Komiyama, Kashikawa, Furusawa,
  Okamura, Graham, Miller, Carter, Mobasher, \& Jogee}]{yagi_dozen_2010}
Yagi, M., Yoshida, M., Komiyama, Y., {et~al.} 2010, The Astronomical Journal,
  140, 1814, \dodoi{10.1088/0004-6256/140/6/1814}

\bibitem[{Zhu {et~al.}(2023)Zhu, Tonnesen, \& Bryan}]{zhu_when_2023}
Zhu, J., Tonnesen, S., \& Bryan, G.~L. 2023, The Astrophysical Journal, 960,
  54, \dodoi{10.3847/1538-4357/acfe6f}

\end{thebibliography}
\bibliographystyle{aasjournal}

\end{document}